\documentclass[aps,prd,showpacs,
twocolumn
,superscriptaddress]{revtex4}
\usepackage{graphicx}
\usepackage{dcolumn}
\usepackage{bm}
\usepackage{color}
\usepackage[normalem]{ulem} 
\usepackage[dvipsnames]{xcolor} 
\usepackage{hyperref}
\hypersetup{
  colorlinks=true,        
  linkcolor=blue,         
  citecolor=cyan,         
}

\usepackage{doi}
\usepackage{amsmath}
\usepackage{amssymb}
\usepackage{commath}

\begin{document}

\title{Dynamics of test particles around\\a Bardeen black hole surrounded by perfect fluid dark matter}

\author{Bakhtiyor Narzilloev}
\email[]{nbakhtiyor18@fudan.edu.cn}
\affiliation{Center for Field Theory and Particle Physics and Department of Physics, Fudan University, 200438 Shanghai, China }
\affiliation{Ulugh Beg Astronomical Institute, Astronomy St. 33, Tashkent 100052, Uzbekistan}

\author{Javlon Rayimbaev}
\email[]{javlon@astrin.uz}
\affiliation{Ulugh Beg Astronomical Institute, Astronomy St. 33, Tashkent 100052, Uzbekistan}
\affiliation{Institute of Nuclear Physics, Ulugbek 1, Tashkent 100214, Uzbekistan}
\affiliation{National University of Uzbekistan, Tashkent 100174, Uzbekistan}

\author{Sanjar Shaymatov}
\email[]{sanjar@astrin.uz}
\affiliation{Institute for Theoretical Physics and Cosmology, Zhejiang University of Technology, Hangzhou 310023, China}
\affiliation{Ulugh Beg Astronomical Institute, Astronomy St. 33, Tashkent 100052, Uzbekistan}
\affiliation{National University of Uzbekistan, Tashkent 100174, Uzbekistan}
\affiliation{Tashkent Institute of Irrigation and Agricultural Mechanization Engineers, Kori Niyoziy, 39, Tashkent 100000, Uzbekistan}

\author{Ahmadjon~Abdujabbarov}
\email[]{ahmadjon@astrin.uz}

\affiliation{Shanghai Astronomical Observatory, 80 Nandan Road, Shanghai 200030, P. R. China}
\affiliation{Ulugh Beg Astronomical Institute, Astronomy St. 33, Tashkent 100052, Uzbekistan}
\affiliation{Institute of Nuclear Physics, Ulugbek 1, Tashkent 100214, Uzbekistan}
\affiliation{National University of Uzbekistan, Tashkent 100174, Uzbekistan}

\affiliation{Tashkent Institute of Irrigation and Agricultural Mechanization Engineers, Kori Niyoziy, 39, Tashkent 100000, Uzbekistan}  

\author{Bobomurat Ahmedov}
\email[]{ahmedov@astrin.uz}
\affiliation{Ulugh Beg Astronomical Institute, Astronomy St. 33, Tashkent 100052, Uzbekistan}
\affiliation{Tashkent Institute of Irrigation and Agricultural Mechanization Engineers, Kori Niyoziy, 39, Tashkent 100000, Uzbekistan}
\affiliation{National University of Uzbekistan, Tashkent 100174, Uzbekistan}

\author{Cosimo Bambi}
\email[]{bambi@fudan.edu.cn}
\affiliation{Center for Field Theory and Particle Physics and Department of Physics, Fudan University, 200438 Shanghai, China }

\date{\today}

\begin{abstract}

  We study the dynamics of (i) neutral test particles, (ii) magnetically charged test particles, and (iii) test magnetic dipole 
   around a regular Bardeen black hole surrounded by perfect fluid dark matter (PFDM). It has been shown how the magnetic charge of the black hole and the parameter of the surrounding PFDM can influence the innermost stable circular orbit (ISCO) radius of a test particle. 
    We have found that the ISCO radius is strongly affected as a consequence of the combined effect of the magnetic charge parameter and the perfect fluid dark matter. The black hole  magnetic charge parameter $g$ and the combined effect of perfect fluid dark matter can mimic the black hole rotation parameter up to $a/M\approx 0.9$. It has been observed that the ISCO for magnetic dipole disappears at the values exceeding the calculated upper value for the magnetic interaction parameter $\beta$. The upper limit decreases with the increase of both the dark matter and magnetic charge parameters. Finally, as an astrophysical application, we have analyzed degeneracy effects of spin of Kerr black holes and magnetic charge of regular Bardeen black holes for the different values of the dark matter parameter providing exactly the same value for ISCO radius of a magnetic dipole with the same value of the parameter $\beta=10.2$ of the magnetar called PSR J1745-2900 orbiting around supermassive black hole Sagittarius A*. It has been observed that the magnetic charge of the pure regular Bardeen black hole can mimic the spin of a Kerr black hole up to $a/M \simeq 0.8085$, while upper limit for the magnetic charge which may provide ISCO for the magnetic dipole is $g_{\rm upper}\simeq 0.65M$. In the presence of PFDM with the parameter $\alpha=0.01 (0.05)$, the upper limit for the magnetic charge decreases and equals to $g_{\rm upper}\simeq 0.62 M$ ($0.548M$) and consequently mimicker value for the spin parameter of black hole lies in the range of $a/M \in (0.0106 \div 0.8231 )$ ( $a/M \in (0.0816 \div 0.8595)$). We also show that the same values of the spin of Kerr black hole and the magnetic charge of regular Bardeen black hole surrounded by PFDM provide the same values for the ISCO radius of the chosen magnetar.

\end{abstract}

\pacs{04.50.-h, 04.40.Dg, 97.60.Gb}

\maketitle

\section{Introduction}

General relativity has been successfully tested in both weak field approximation using solar system tests and strong field regime using observation of the shadow of the supermassive black hole (SMBH) at the center of elliptical galaxy M87*~\cite{EHT19a,EHT19b} by Event Horizon Telescope and direct detection of gravitational waves by the LIGO-VIRGO Collaboration~\cite{LIGO16a,LIGO16} from the coalescence of black holes (BHs) and neutron stars close binaries. However, general relativity meets a fundamental problem connected with the exact solutions of field equations containing physical singularity. It is expected that the singularity problem will be resolved with construction of quantum gravity theory. However, Bardeen had obtained the singularity free black hole solution within the general relativity coupled with special case of nonlinear electrodynamics~\cite{Bardeen68}. Other regular black hole solutions have been obtained by Hayward~\cite{Hayward06}, and Ayon-Beato and Garcia~\cite{Ayon-Beato98,Ayon-Beato99,Ayon-Beato99a}. The rotating analogue of these solutions has been obtained in~\cite{Bambi13,Toshmatov14,Toshmatov17d,Toshmatov18b}. The properties of regular black hole solutions coupled with nonlinear electrodynamics have been studied in~Refs.~\cite{Toshmatov15b,Toshmatov15a,Dymnikova15}.


The existence of dark energy and dark matter in the Universe has been independently proven by several independent experiments and observations~\cite{Planck16}. This of course motivates researchers to consider black holes surrounded by dark matter. Particularly, Kiselev has obtained a black hole solution surrounded by quintessence~\cite{Kiselev03} and its rotational analogue obtained in~\cite{Toshmatov17}.  The properties of black holes surrounded by quintessence have been studied in~\cite{Abdujabbarov17b,Shaymatov18,Chakrabarty19,Benavides20}. Authors of Refs.~\cite{Ghosh18,Zhang06,Ghosh16a} have studied the properties of the  black holes surrounded by quintessence.
{Kiselev proposed another possible model for dark matter for which a static and spherically symmetric black hole solution with a dark matter fluid as a quintessential scalar field has been derived~\cite{Kiselev03}. This solution includes a logarithmic term $\alpha\ln(r/r_q)$ appearing in the metric functions, corresponding to the non-vanishing contribution of dark matter. It is well known that the standard assumption implies that the dark mater halo is formed from the weakly interacting massive particles satisfying $\omega\simeq 0$ for equation of state. Later on, following this assumption, Li and Yang~\cite{Li12} derived a similar black hole solution involving a term $(r_q/r)\ln(r/r_q)$, suggesting the dark matter halo as a phantom scalar field (i.e. where $r=r_q$ implies the boundary of the dark matter halo). In this model developed by Li and Yang~\cite{Li12}, the phantom scalar field does not act as a cosmological role, thus leading to realize  the distribution of background matter as a dark matter.  }
In Refs.~\cite{Haroon19,Hou18,Shaymatov20d}, other astrophysically relevant properties of black holes surrounded by perfect fluid dark matter have been considered.


The electromagnetic field around an astrophysical compact object is an essential part of its properties related to astronomical observations and affects the dynamics of charged and magnetic dipole. 
Despite the no-hair theorem, according to which a black hole cannot have its own magnetic field~\cite{Misner73}, one may consider black holes immersed in an external magnetic field. Particularly, the solution of Maxwell equations around a black hole immersed in an external asymptotically uniform magnetic field has been obtained in~\cite{Wald74}. The magnetic field changes its structure due to curved spacetime and affects the dynamics of charged particle motion~ \cite{Chen16,Hashimoto17,Dalui19,Han08,Moura00,
Morozova14}.
The study of the structure of the spacetime and test particle motion around black holes can be found in Refs.~\cite{Jawad16,Hussain15,Jamil15,
Hussain17,Babar16,Banados09,Majeed17,
Zakria15,Brevik19,DeLaurentis2018PhRvD, Shaymatov13, Atamurotov13a,Narzilloev20a}. A number of studies are devoted to study the electromagnetic field  and particle motion around black holes in the presence of an external magnetic field~\cite{Kolos17,Kovar10,Kovar14,Aliev89,Aliev02,
Aliev86,Frolov11,Frolov12,Stuchlik14a,Shaymatov14,
Abdujabbarov10,Abdujabbarov11a,Abdujabbarov11,
Abdujabbarov08,Karas12a,Shaymatov15,Stuchlik16,
Rayimbaev20,Turimov18b,Turimov17,
Shaymatov20egb,Rayimbaev15,shaymatov19b,Rayimbaev19,
Shaymatov20b,Narzilloev2020C,Narzilloev19}.
Other possible way of testing the spacetime and gravity theory in strong field regime is to study the magnetic dipole dynamics around compact object in the presence of magnetic field. One of the attempts to study and analyze the magnetic dipole motion around black hole  in the presence of external asymptotically uniform magnetic field has been made in~\cite{deFelice,defelice2004}. 
The magnetic dipole motion around non-Schwarzschild black holes is considered in~\cite{Rayimbaev16} in the presence of magnetic field. Collisions of such particles near rotating black holes in quintessence have been investigated in~\cite{Oteev16}. We also refer the reader to the works where acceleration of magnetic dipole near gravitational compact objects has been studied in the presence of external magnetic field~\cite{Toshmatov15d,Abdujabbarov14,
Rahimov11a,Rahimov11,Haydarov20,Haydarov20b}.

Here we plan to study the dynamics of neutral, magnetically charged, and magnetic dipole around a Bardeen regular black hole surrounded by perfect fluid dark matter. The paper is organized as follows: 
in Sect.~\ref{sec2}  we have studied test particle motion around a Bardeen black hole surrounded by perfect fluid dark matter. Sect.~\ref{Sec:motion} is devoted to the dynamics of magnetically charged particles around regular black holes surrounded by perfect fluid dark matter. We explore the magnetic dipole motion around a Bardeen regular black hole in the presence of perfect fluid dark matter in Sect.~\ref{section3}. Sect.~\ref{application} is devoted to possible astrophysical applications of the obtained results. We conclude and summarize our results in Sect.~\ref{conclusion}.
  We use geometrized units $G=c=1$, $(-,+,+,+)$ is the signature of the spacetime, and Greek (Latin) indices run 0,1,2,3 (1,2,3).

 \section{Test particle motion around Bardeen BH\label{sec2}}

The spacetime metric for a central object described by Bardeen in the perfect fluid dark matter (PFDM) has the following form~\cite{2007.09408}
\begin{eqnarray}\label{metric}
ds^2&=& -f(r) dt^2+f(r)^{-1} dr^2+r^2 d\Omega^2 \ ,
\\\nonumber 
f(r)&=&1-\frac{2 M r^2}{(r^2+g^2)^{\frac{3}{2}}} +\frac{\alpha}{r} \ln{\frac{r}{|\alpha|}}\ ,
\end{eqnarray}
with $g$ being the magnetic charge, $M$ being black hole mass, and $\alpha$ related to the perfect fluid dark matter density and pressure. In the case of vanishing $g$ and $\alpha$, the spacetime metric (\ref{metric}) reduces to the Schwarzschild metric. For the given spacetime metric in the case of nonvanishing $\alpha\neq 0$, the stress energy-momentum tensor of the perfect fluid dark matter distribution is given by 
\begin{eqnarray}\label{Eq:T}
T^\mu_\nu={\rm diag}(-\rho,p_{r},p_{\theta},p_{\phi})\, ,
\end{eqnarray}
with the density, radial and tangential pressures 
\begin{eqnarray}\label{Eq:rho}
\rho=-p_{r}= \frac{\alpha}{8\pi r^3}\,  \mbox{~~~and~~~} p_{\theta}=p_{\phi}=\frac{\alpha}{16\pi r^3}\, .
\end{eqnarray}
Note that for a perfect fluid dark matter distribution we shall restrict ourselves to the case $\alpha>0$.

\begin{figure*}
	\begin{center}
		a.
		\includegraphics[width=0.45\linewidth]{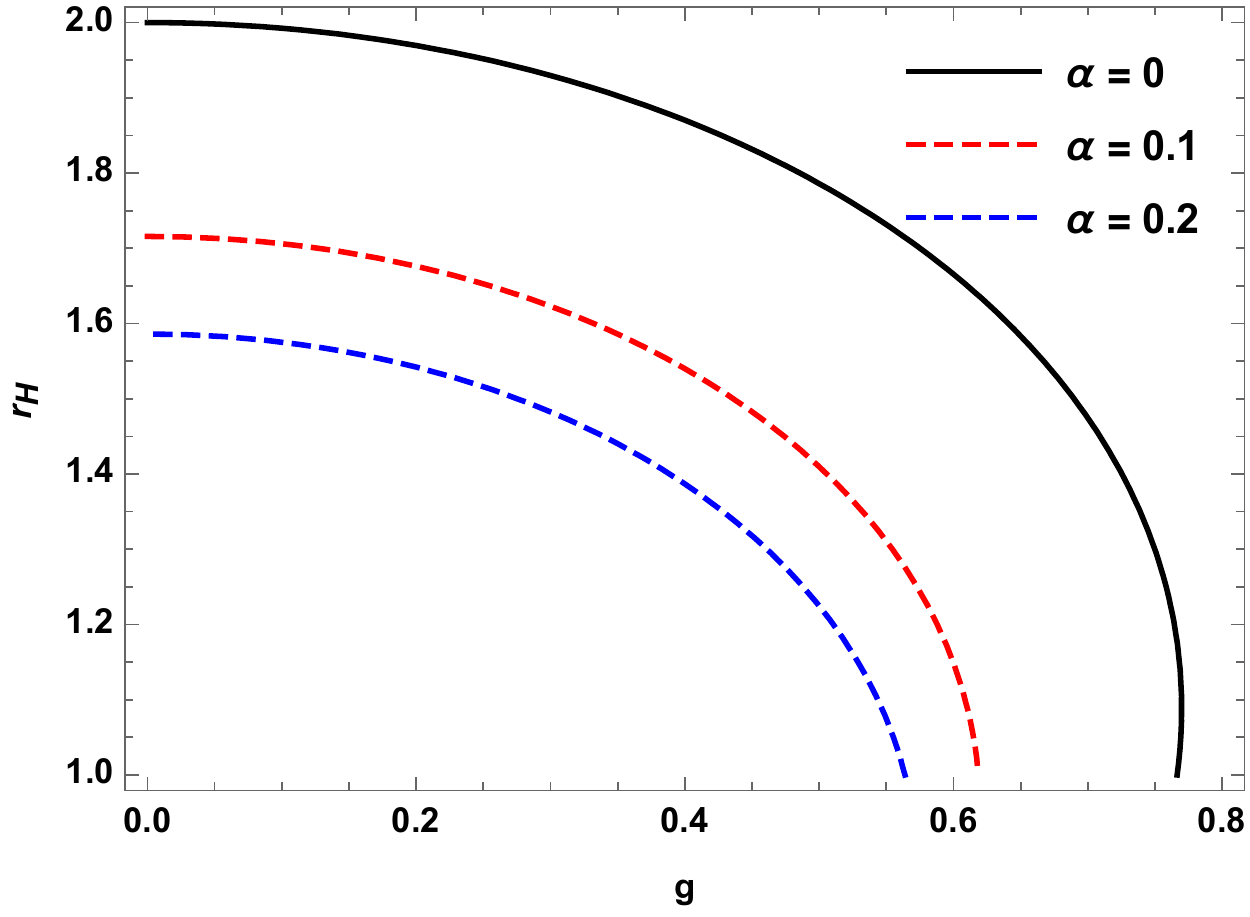}
		b.
		\includegraphics[width=0.45\linewidth]{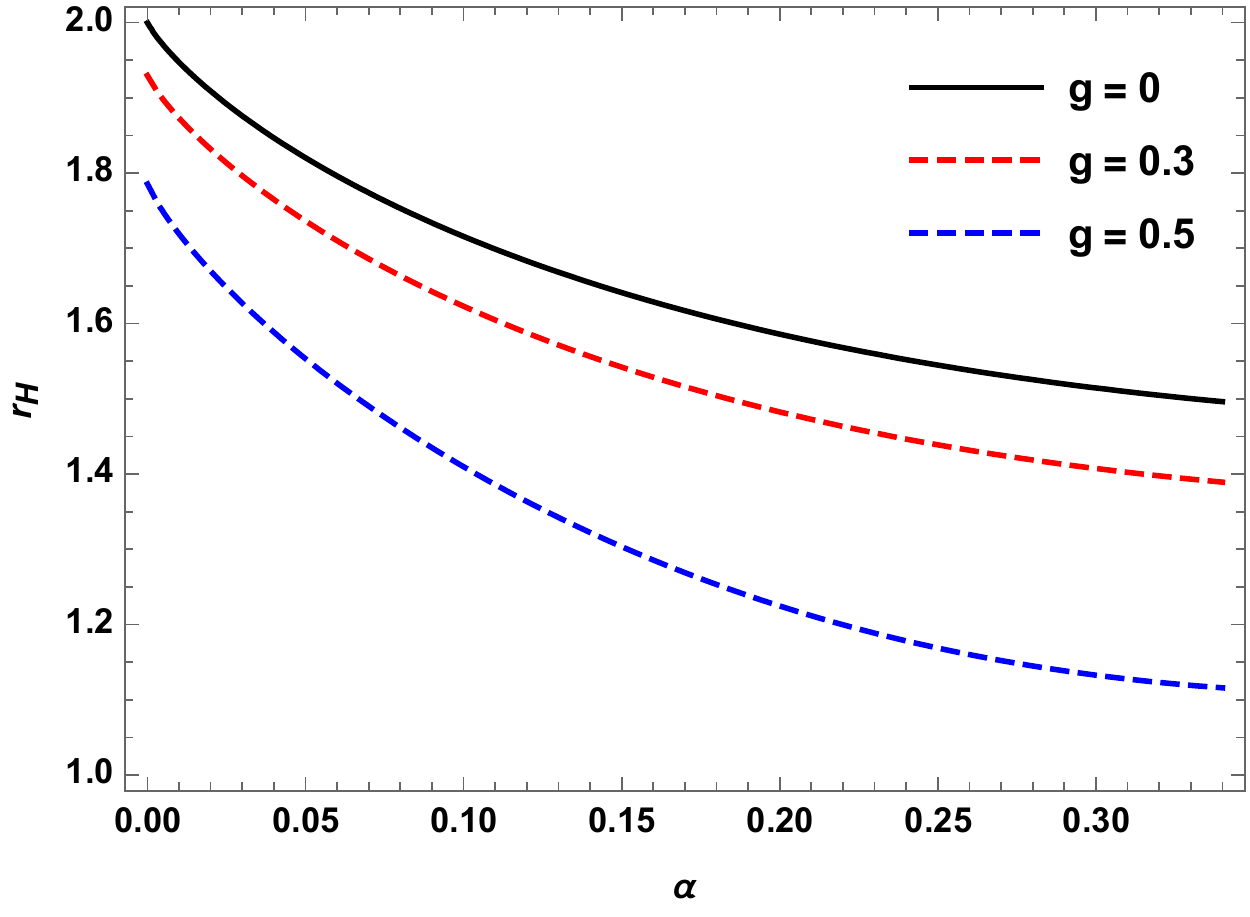}
	\end{center}
	\caption{Dependence of event horizon radius on the parameters $g$ and $\alpha$. Left (right) panel is for dependence of $r_H$ from $g$ ($\alpha$) for fixed values of $\alpha$ ($g$). \label{rh}}
\end{figure*}

 The event horizon of the black hole is located at the real root of $g^{rr}=0$, and its dependence on black hole parameters is shown in Fig.~\ref{rh}. Whether it is the spacetime singularity or not, it is worth calculating Kretschman scalar, i.e., 
 $\mathcal{K}=R_{\alpha\beta\mu\nu}R^{\alpha\beta\mu\nu}$, which reads as 

\begin{eqnarray}
\mathcal{K}&=&\frac{12 \alpha^2 \ln^2\frac{r}{\left| \alpha\right| }}{r^6}+\frac{13 \alpha^2}{r^6}
\\\nonumber
&+&\frac{4 \alpha \ln\frac{r}{\left|
		\alpha\right| } \left[\frac{6 M r^5 \left(3 g^2-2 r^2\right)}{\left(g^2+r^2\right)^{7/2}}-5 \alpha\right]}{r^6}
	\\\nonumber
	&+&\frac{4 \alpha M \left(-2
	g^4-37 g^2 r^2+10 r^4\right)}{r^3 \left(g^2+r^2\right)^{7/2}}
\\\nonumber
&+&\frac{12 M^2 \left(8 g^8-4 g^6 r^2+47 g^4 r^4-12 g^2 r^6+4 r^8\right)}{\left(g^2+r^2\right)^7}\ .
\end{eqnarray}

The real singularity appears only at $r=0$ as seen in the above equation.

 One can observe that that an increase in the magnetic charge parameter $g$ decreases the event horizon radius, thus leading to the assumption that its behavior is similar to the rotation parameter of Kerr spacetime metric, which also decreases the black hole horizon.  
 Let us now turn to the charged particle motion around the black hole. 
 In doing so, we use the well-known Hamilton-Jacobi equation of motion, and for the charged particles it reads as

\begin{eqnarray}\label{HJ}
g^{\alpha\beta} \left(\frac{\partial S}{dx^\alpha}+q A_\alpha \right) \left(\frac{\partial S}{dx^\beta}+q A_\beta \right)=-m^2\ .
\end{eqnarray}

Here $A_\alpha$ defines the four vector potential of electromagnetic field around the black hole generated by the magnetic charge of a latter one. This $U(1)$ gauge field has the following form (see~[2007.09408])

\begin{eqnarray}
A_\alpha=-\delta^\phi_\alpha g \cos\theta\ ,
\end{eqnarray}
where $\delta^\mu_\nu$ denotes the Kronecker delta. Since we are mostly interested in the equatorial motion of a test particle it is obvious that given gauge field becomes zero on that plane ($\theta=\pi/2$). Thus, in this section, we reduce our calculations to the investigation of only neutral test particles ($q=0$) for which the equation of motion (\ref{HJ}) reduces to

\begin{eqnarray}\label{HJ2}
g^{\alpha\beta} \frac{\partial {\cal S}}{dx^\alpha} \frac{\partial {\cal S}}{dx^\beta}=-m^2\ .
\end{eqnarray}

One can write the action for a test particle as

\begin{eqnarray}\label{action1}
{\cal S}=-E t + L \phi + {\cal S}_r + {\cal S}_\theta\ ,
\end{eqnarray}
which comes from the symmetry of a system. Here $E$ and $L$ define the energy and angular momentum of a test particle, respectively. The functions ${\cal S}_{r}$ and ${\cal S}_\theta$  are functions of $r$ and $\theta$, respectively. Then the equation of motion reads

\begin{eqnarray}\label{eom}
&&-\frac{E^2}{1-\frac{2 M
		r^2}{\left(g^2+r^2\right)^{3/2}}+\frac{\alpha}{r} \ln\frac{r}{|\alpha|}}
	\\\nonumber
	&&+\left[1-\frac{2 M
	r^2}{\left(g^2+r^2\right)^{3/2}}+\frac{\alpha}{r} \ln\frac{r}{|\alpha|}\right]  \left(\frac{\partial {\cal S}}{dr}\right)^2
\\\nonumber
&&+\frac{L^2 \csc (\theta
	)}{r^2}+\frac{1}{r^2} \left(\frac{\partial {\cal S}}{d\theta}\right)^2=-m^2\ .
\end{eqnarray}

\begin{figure*}
	\centering
	\includegraphics[width=0.95\linewidth]{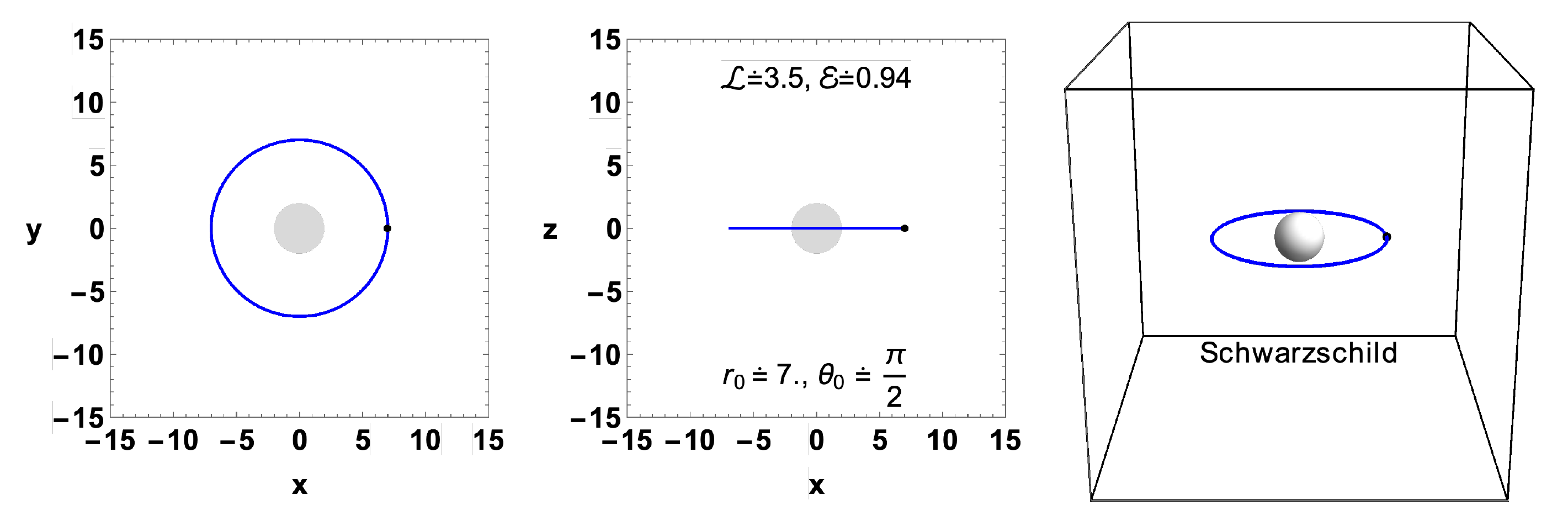}
	\includegraphics[width=0.95\linewidth]{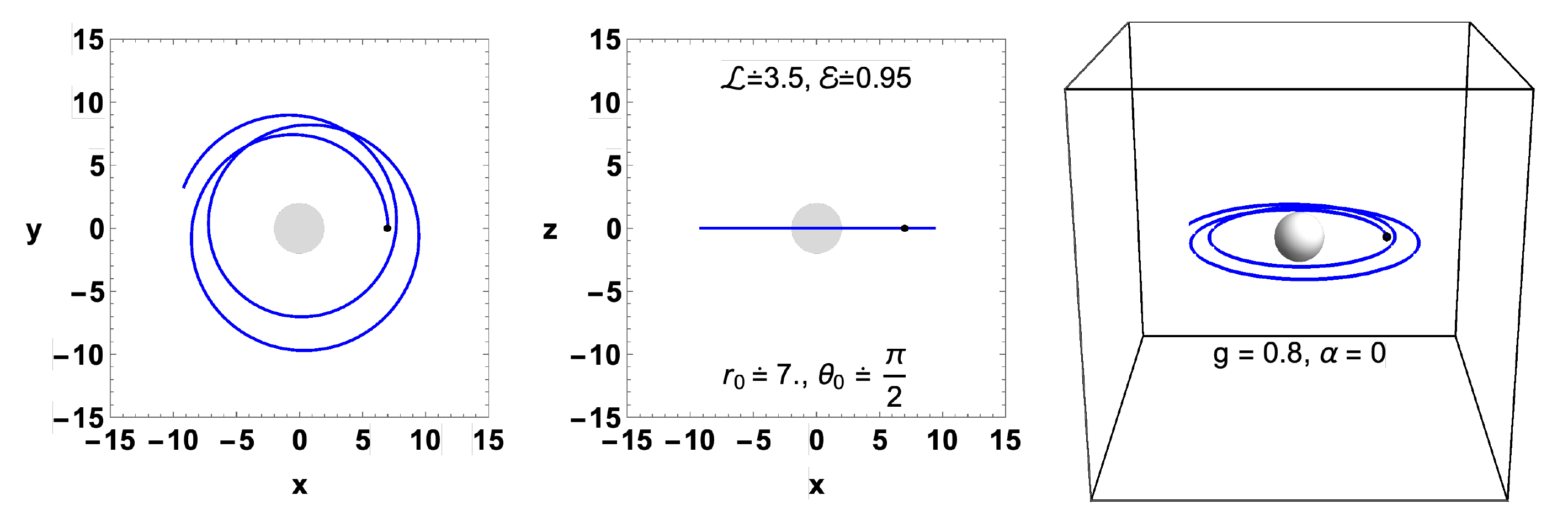}
	\includegraphics[width=0.95\linewidth]{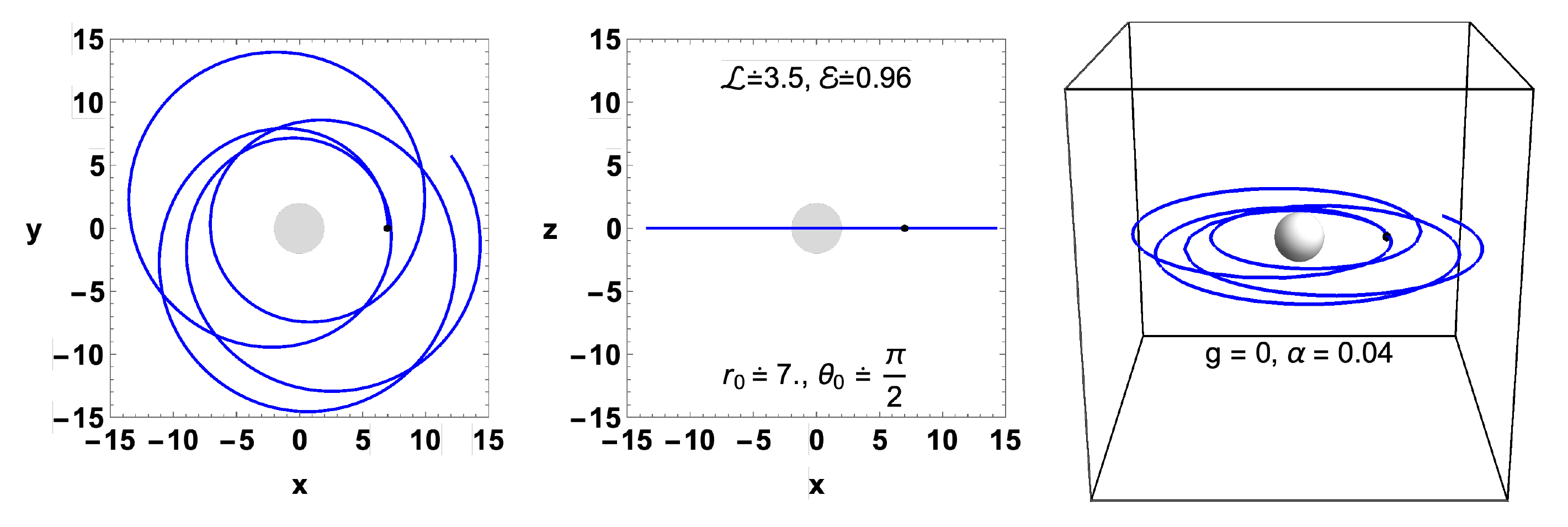}
	\caption{Test particle trajectories for fixed values of $g$ and $\alpha$ parameters. The first row is responsible for the Schwarzschild black hole. The second row is for magnetically charged black hole case. For the third line, the parameter $\alpha$ is nonzero. }
	\label{trj}
\end{figure*}

Trajectory that comes from the equation of motion (when $\theta=\pi/2$) is plotted in Fig. ~\ref{trj}. One can see from the trajectories of a neutral test particle that for a given angular momentum of a particle the presence of magnetic charge $g$ and intensity of PFDM, $\alpha$, makes the average orbit radius of a test particle bigger which means that increase of these parameters makes the resultant interaction between particle and black hole weaker, again similarly to the case of the spin parameter $a$, of the Kerr metric.

Once the equation of motion is obtained, one can write the effective potential of a test particle on equatorial plane ($\theta=\pi/2$) as

\begin{eqnarray}\label{V}
V_{\rm eff}=\left[1-\frac{2 M r^2}{\left(g^2+r^2\right)^{3/2}}+\frac{\alpha}{r} \ln \frac{r}{|\alpha|}\right]\left(1+\frac{L^2}{r^2}\right) 
\end{eqnarray}

\begin{figure*}[t!]
	\begin{center}
		a.
		\includegraphics[width=0.45\linewidth]{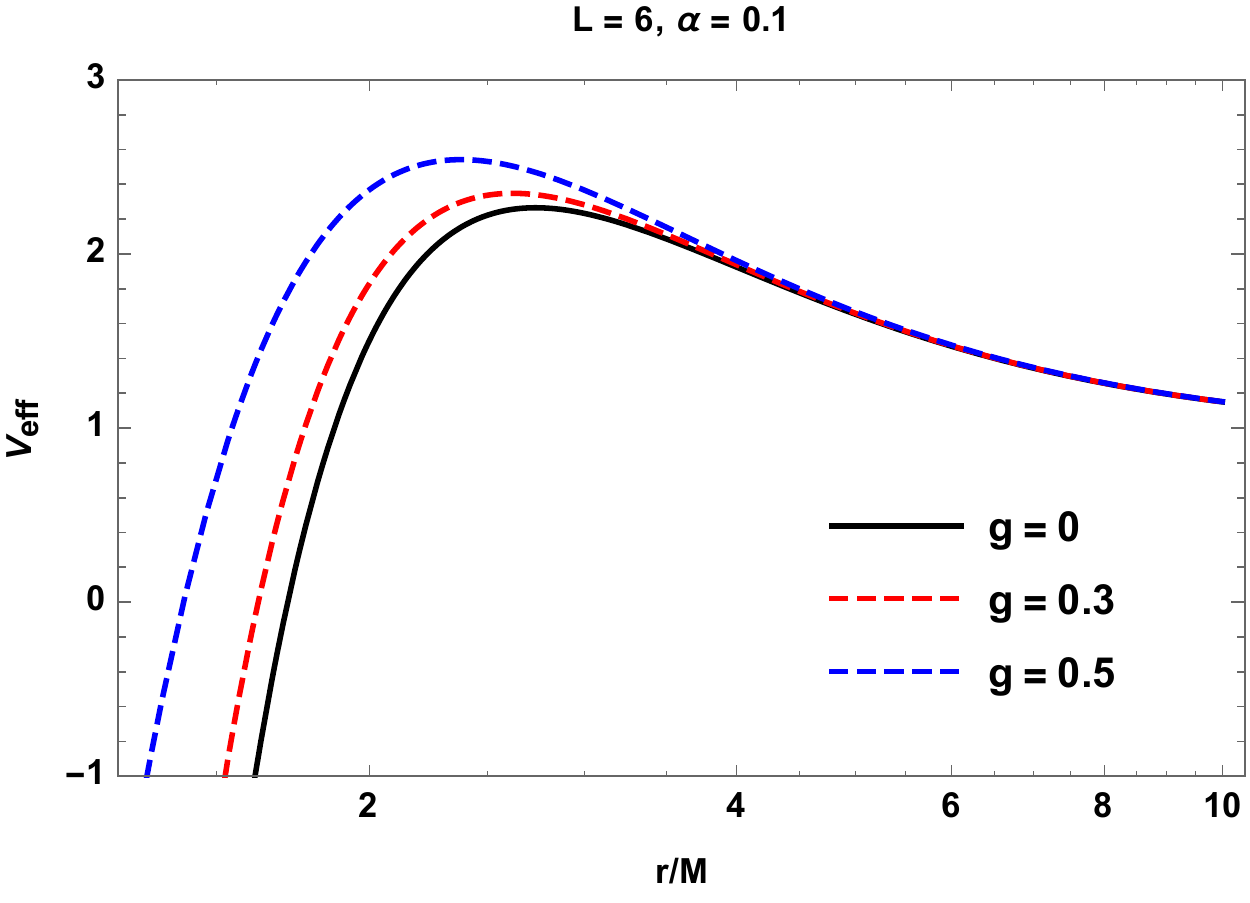}
		b.
		\includegraphics[width=0.45\linewidth]{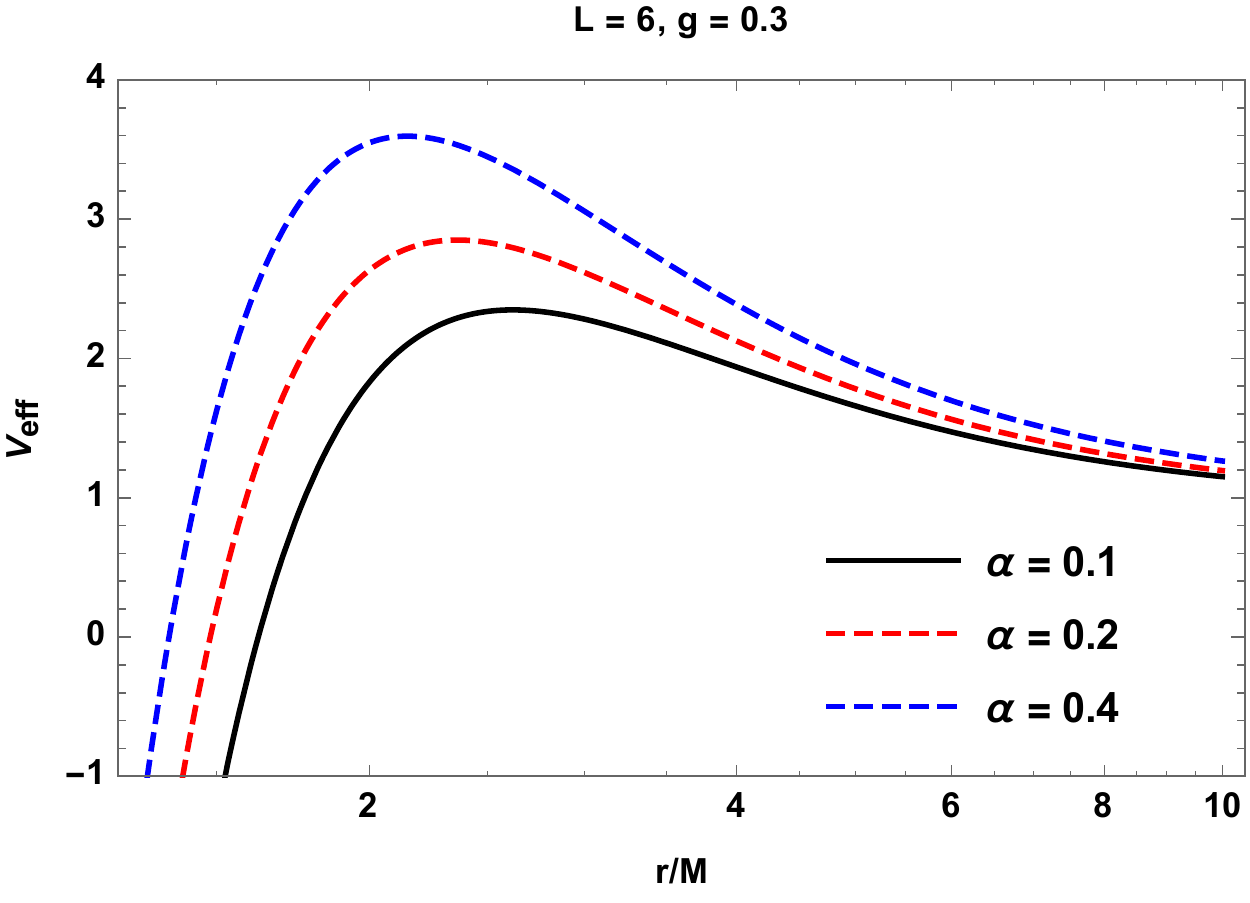}
	\end{center}
	\caption{The radial dependence of the effective potential of
		a test particle around static, magnetically charged black
		hole in PFDM.  \label{Veff}}
\end{figure*}

The radial dependence of the effective potential for fixed values of magnetic charge parameter $g$ and $\alpha$, is plotted in Fig. ~\ref{Veff}.

\begin{figure*}[t!]
	\begin{center}
		a.
		\includegraphics[width=0.45\linewidth]{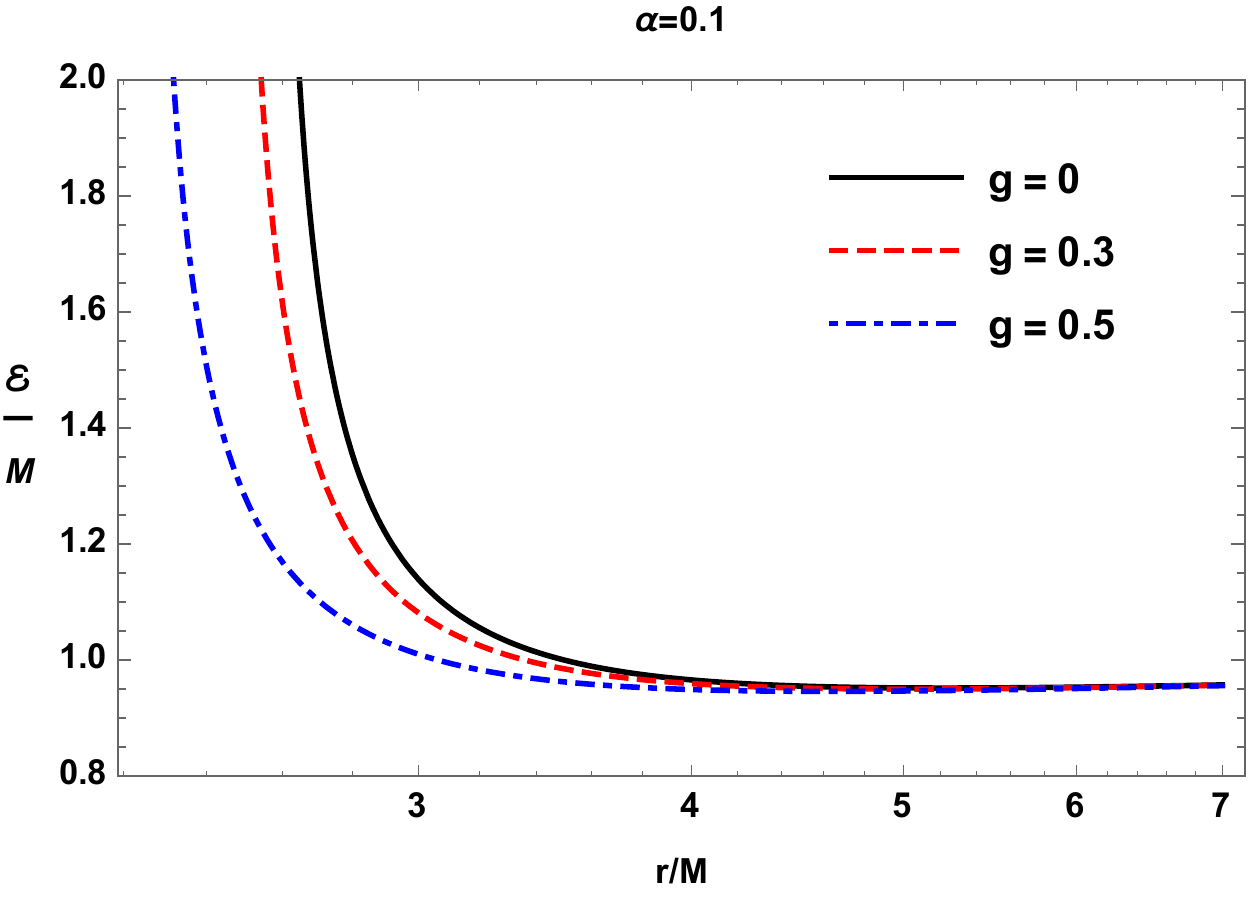}
		b.
		\includegraphics[width=0.45\linewidth]{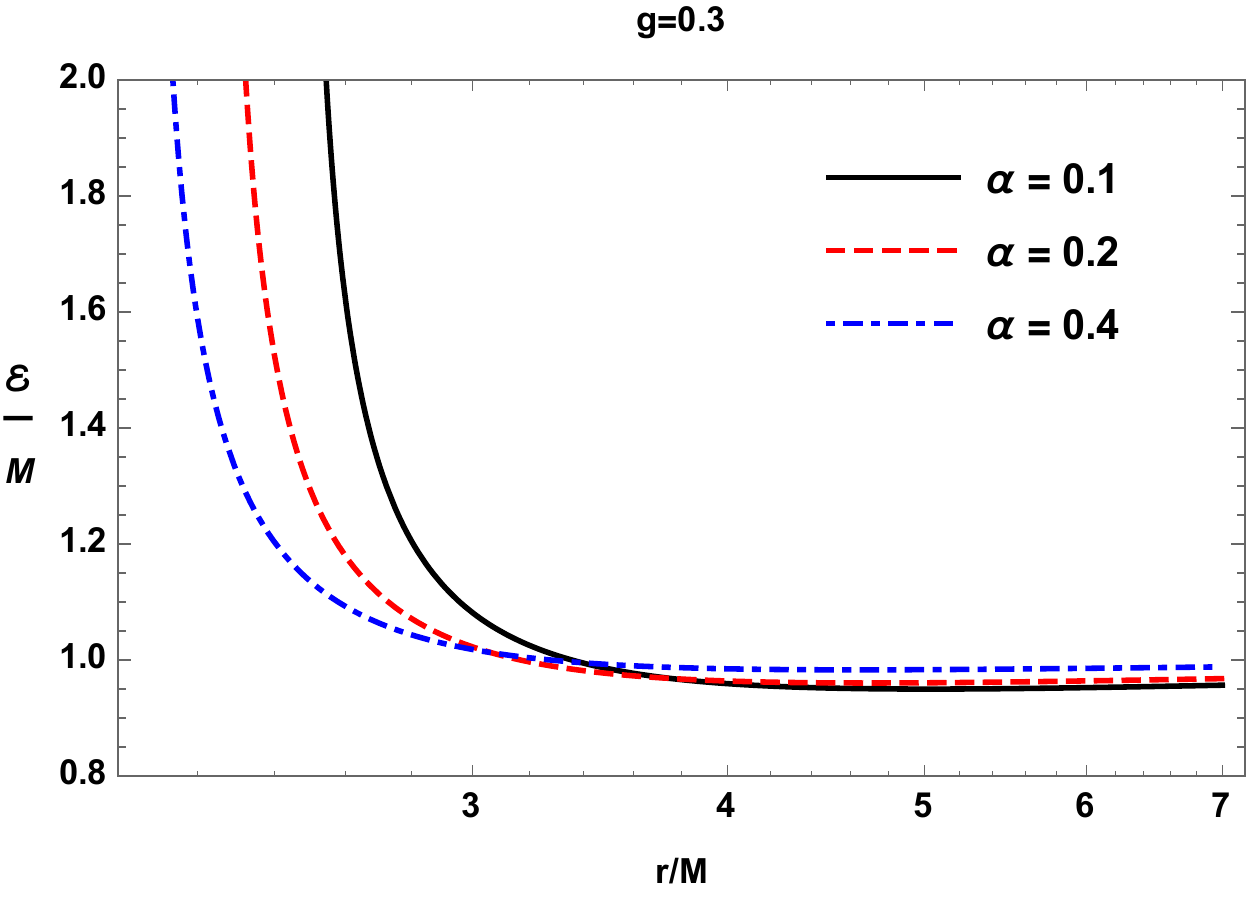}
	\end{center}
	\caption{The radial dependence of the energy of a test particle on circular orbit radius.  The left panel is for black hole with magnetic charge. The right panel is for black hole in PFDM. \label{E}}
\end{figure*}

For the particle to have circular orbits on equatorial plane, the effective potential should satisfy the following set of equations

\begin{eqnarray}\label{c1}
V_{\rm eff}(r)=\mathcal{E}\ ,  V_{\rm eff}'(r)=0\ ,
\end{eqnarray}
which leads the energy of a test particle to have the radial dependence as presented in Fig. ~\ref{E}.

\begin{figure*}[t!]
	\begin{center}
		a.
		\includegraphics[width=0.45\linewidth]{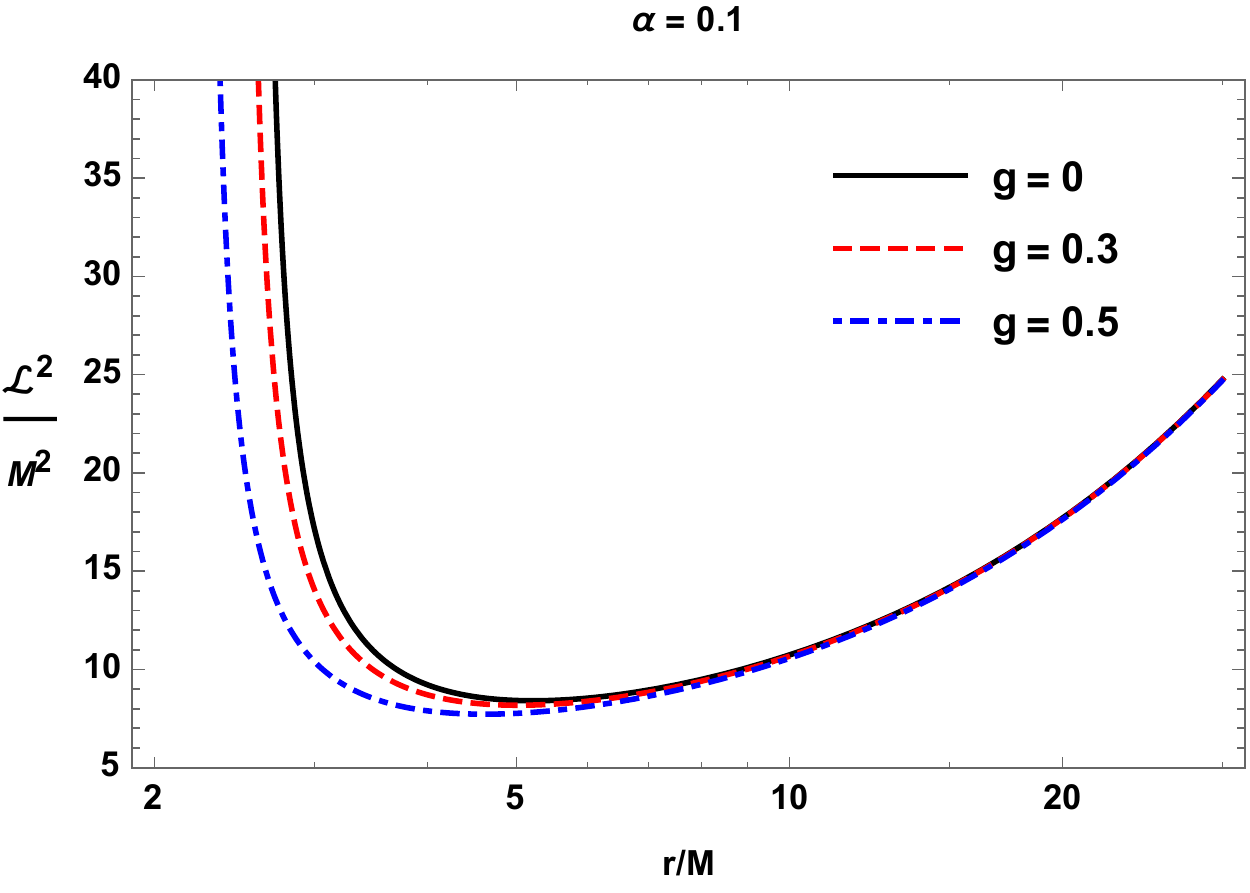}
		b.
		\includegraphics[width=0.45\linewidth]{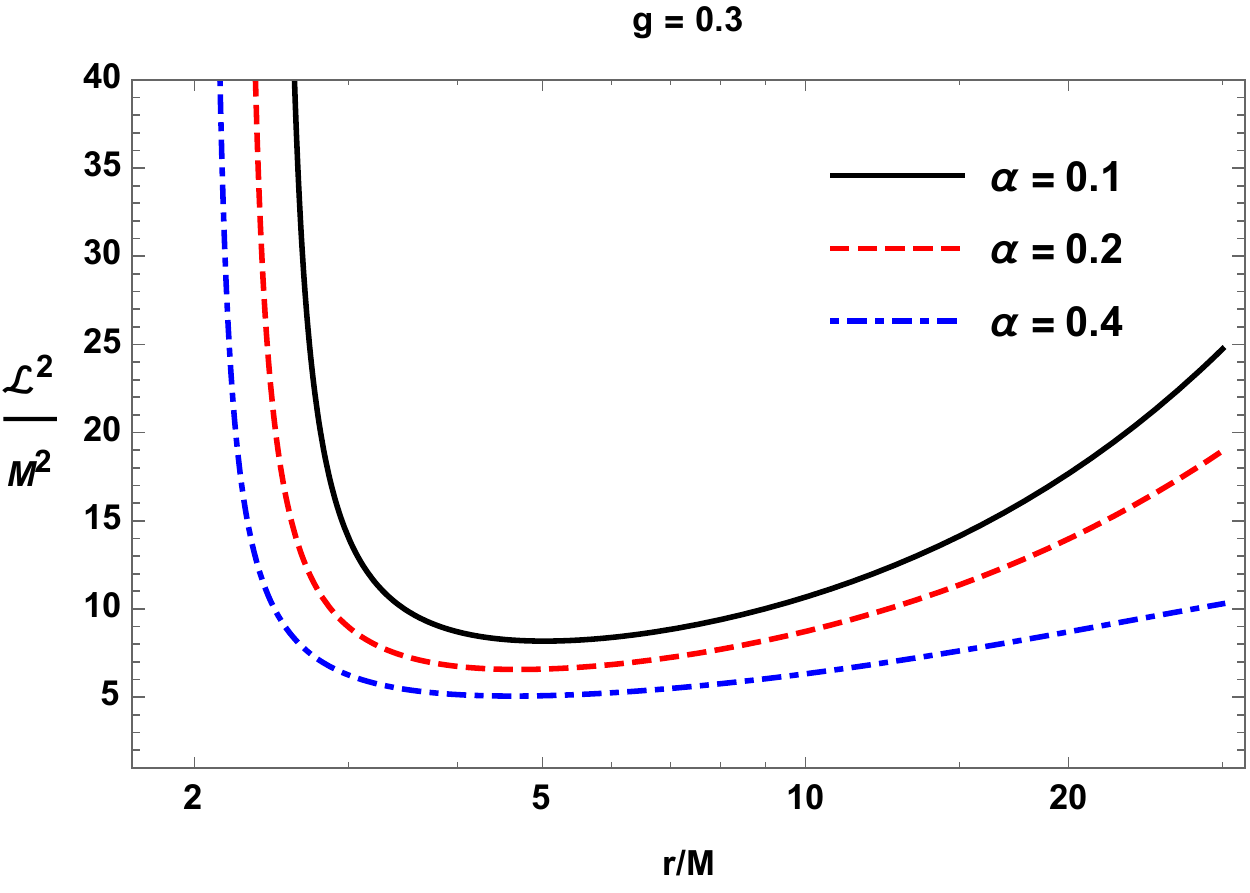}
	\end{center}
	\caption{The radial dependence of the angular momentum of a particle on circular orbit radius. The left panel is for black hole with magnetic charge. The right panel is for black hole in PFDM.  \label{L}}
\end{figure*}

Based on the conditions above one can also plot the radial dependence for angular momentum of a test particle as presented  in Fig. ~\ref{L}. Since the minimum of the angular momentum provides the values for the innermost stable circular orbit (ISCO) radius of a test particle for the given spacetime metric one can ensure for one more time from Fig. ~\ref{L} that the increase of the magnetic charge of a black hole and intensity of PFDM reduces the ISCO radius which means that the interaction between test particle and a black hole becomes weaker. { Indeed, when the attractive force becomes weaker the ISCO radius decreases since it defines the last stable circular orbits. In other words, for weaker attractive force from the central object, particles can orbit more close to the central mass not falling into it which in turn defines the ISCO. In the Kerr spacetime, one can see the same analogy where increase of the rotation parameter from $0$ to $a\rightarrow1$ decreases the ISCO radius from $6M$ to $M$, respectively, which is related with the weakening of the attractive force.}

We have discussed in detail  the similarity between magnetic charge of a Bardeen's black hole together with intensity of PFDM and the spin of a Kerr black hole. Now, it is time to answer to the following question: whether the magnetic charge of the Bardeen's black hole together with intensity of PFDM has the similarity with the spin parameter of a Kerr black hole or is it possible to say that black holes in the Universe might be magnetized static ones surrounded with the dark matter around rather than spinning Kerr ones? We try to answer this question from the idea that if the magnetic charge of a static black hole and PFDM with the metric (\ref{metric}) can somehow mimic the rotation parameter of a Kerr one, then we should have some degeneracy between these parameters providing the same ISCO radii. Based on this idea, let us determine the ISCO radius of a test particle in the spacetime (\ref{metric}) of the magnetized static black hole surrounded with PFDM. One can do this adding auxiliary condition on (\ref{c1}) as
\begin{eqnarray}\label{isco_eqn}\nonumber
V_{\rm eff}''(r)&=&0\ .
\end{eqnarray}

\begin{figure*}[t!]
	\begin{center}
		a.
		\includegraphics[width=0.45\linewidth]{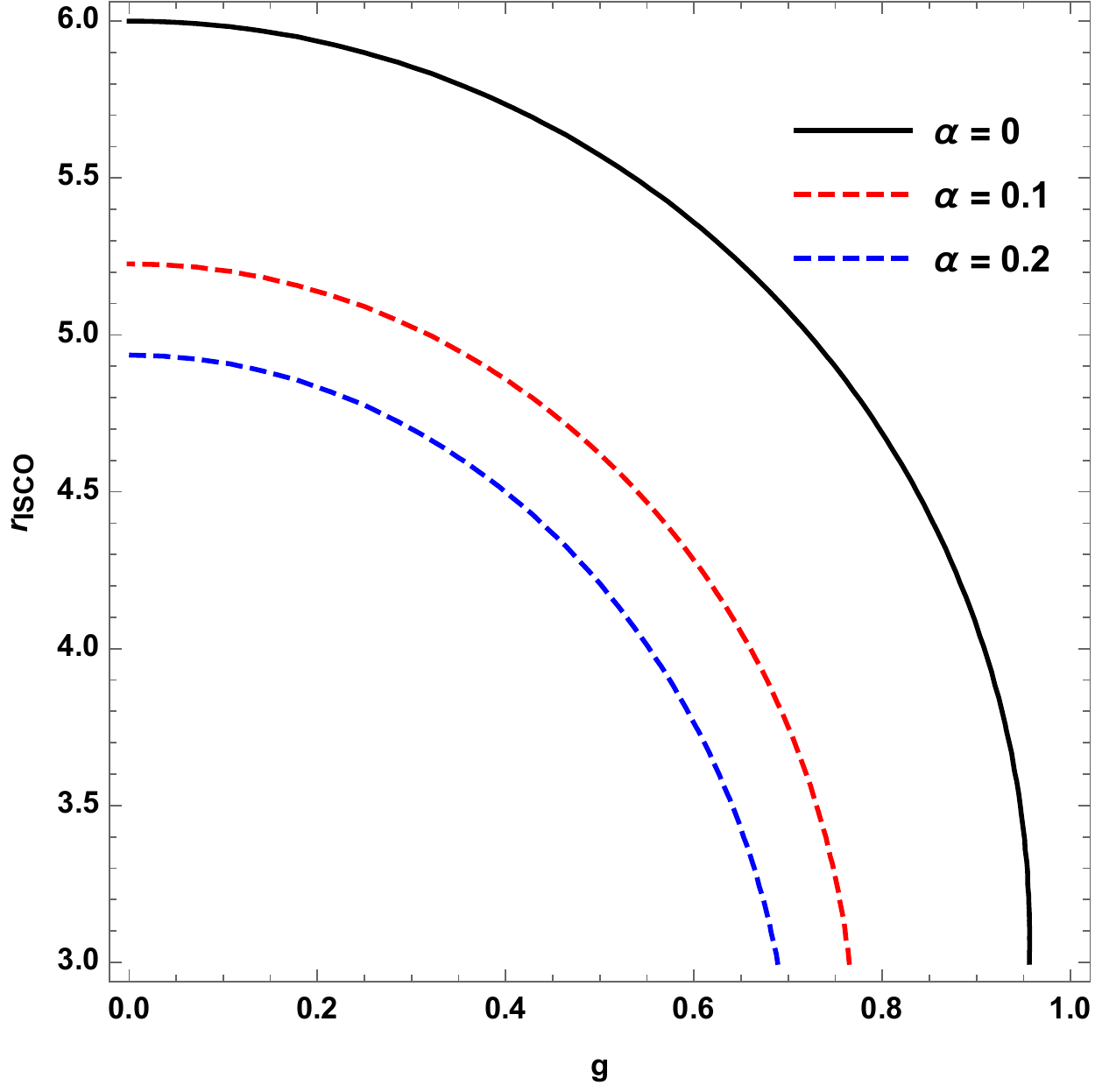}
		b.
		\includegraphics[width=0.45\linewidth]{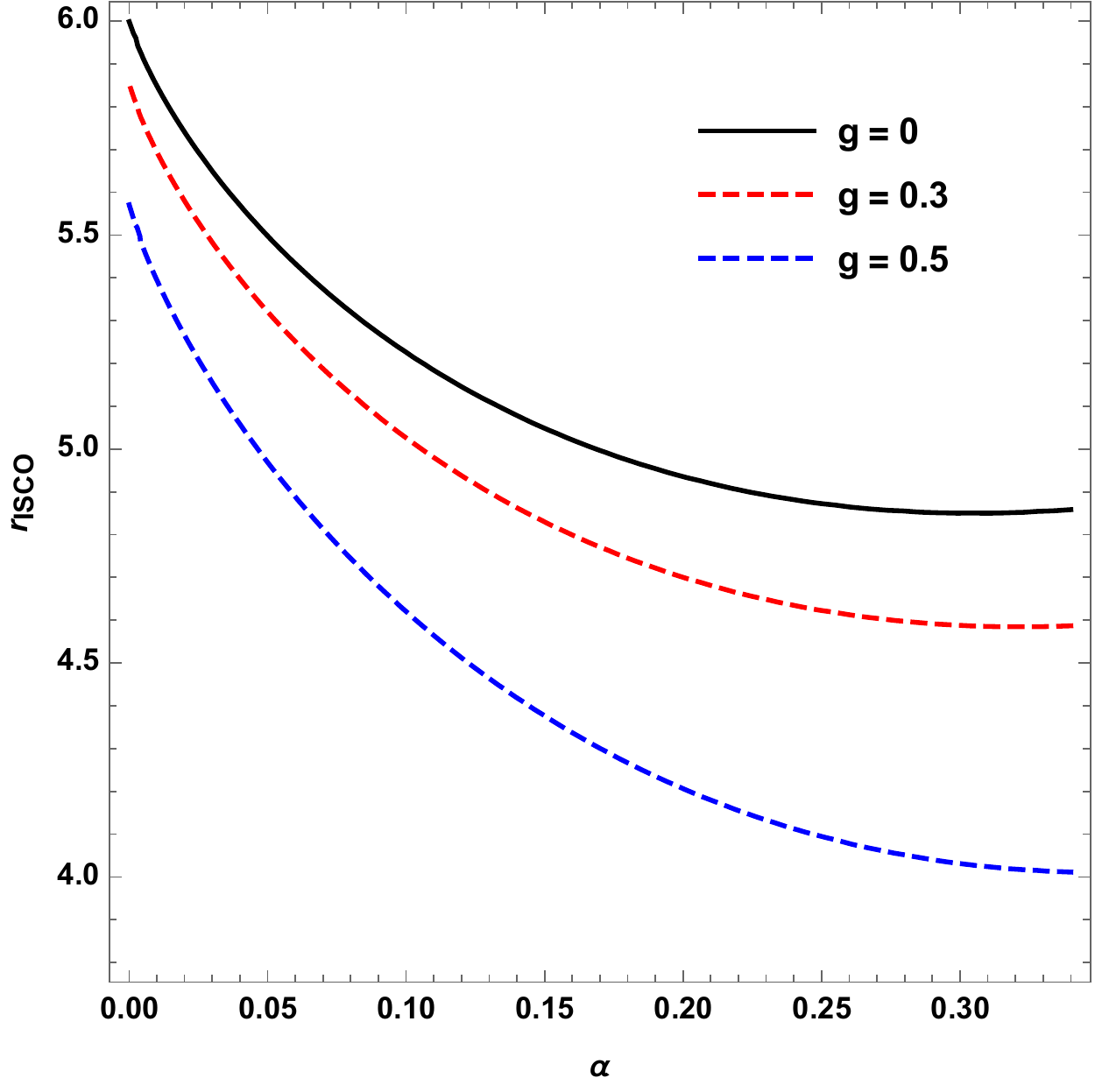}
	\end{center}
	\caption{The dependence of ISCO radius from magnetic charge of a black hole $g$ (left panel, for the fixed values of $\alpha$) and parameter  $\alpha$ (right panel, for the fixed values of magnetic charge $g$).  \label{isco}}
\end{figure*}

From these set of conditions, one can plot the dependence between ISCO radius and magnetic charge parameter of a black hole and intensity of PFDM as in Fig. ~\ref{isco}. We see that the magnetic charge parameter can reduce the ISCO radius up to $3 M$ and the intensity of PFDM up to $5 M$ (when $g=0$) while the rotation parameter of a Kerr black hole can do this task better (up to $M$).

\begin{figure*}[t!]
	\begin{center}
		a.
		\includegraphics[width=0.3\linewidth]{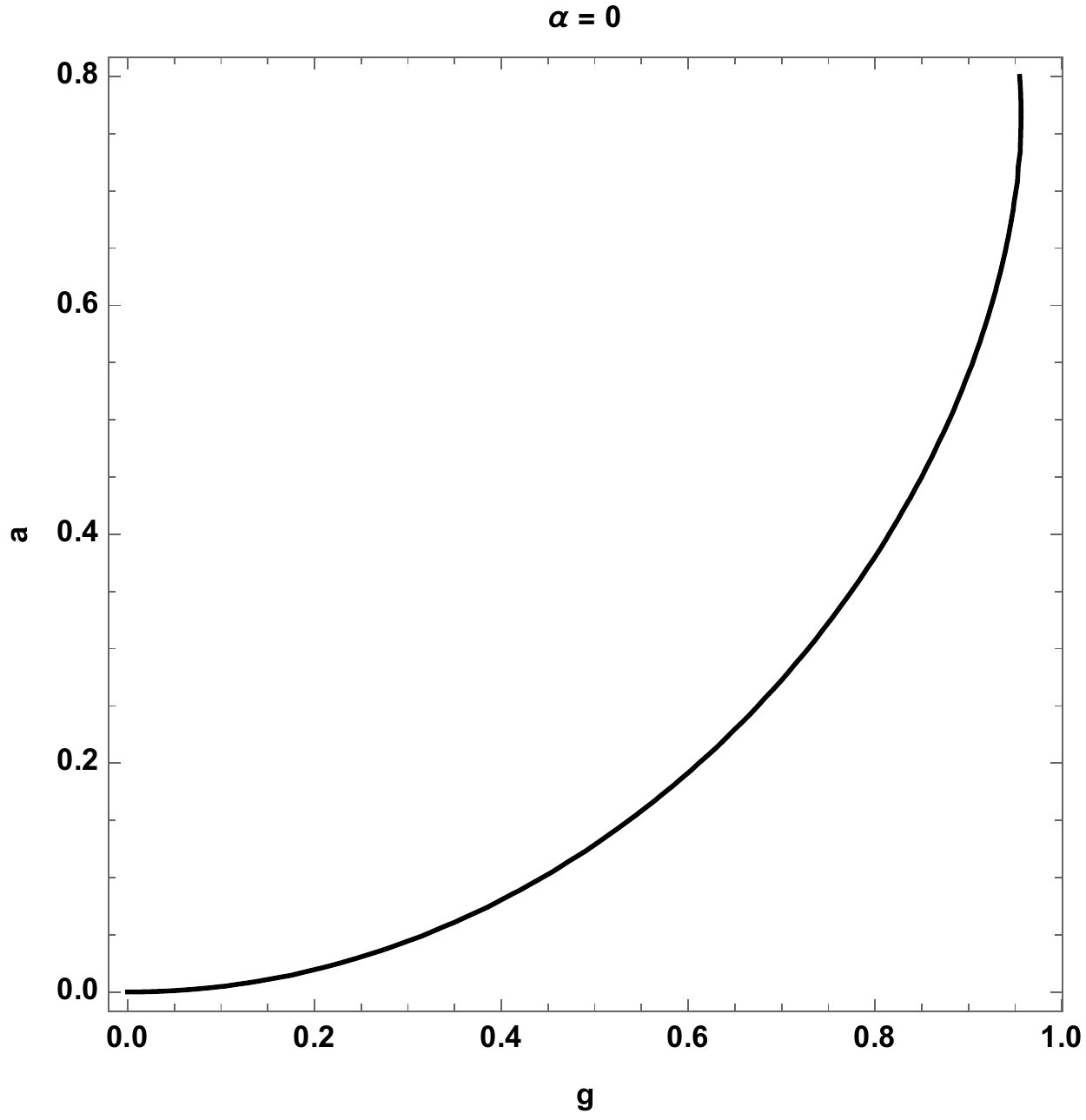}
		b.
		\includegraphics[width=0.3\linewidth]{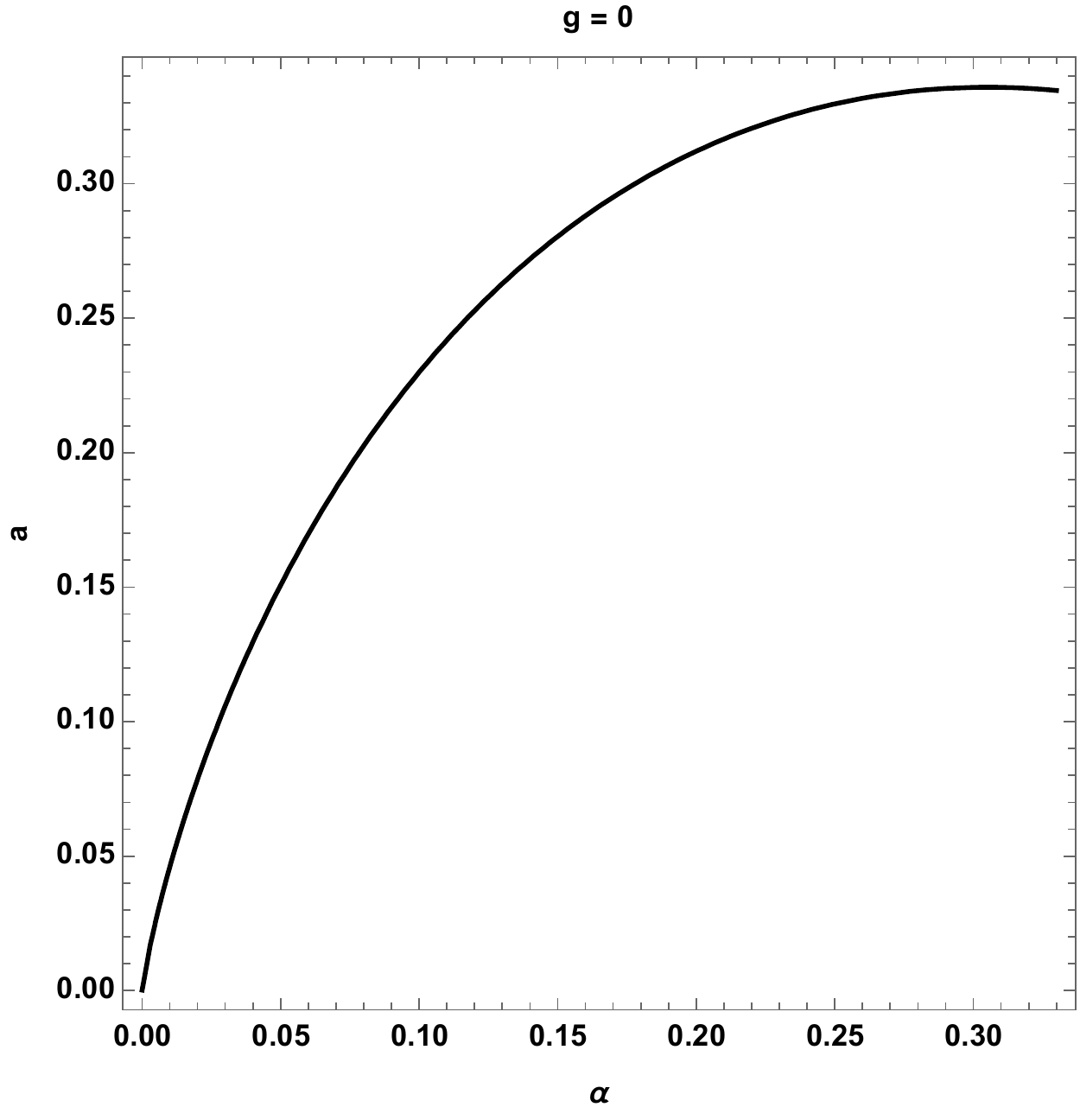}
		c.
		\includegraphics[width=0.3\linewidth]{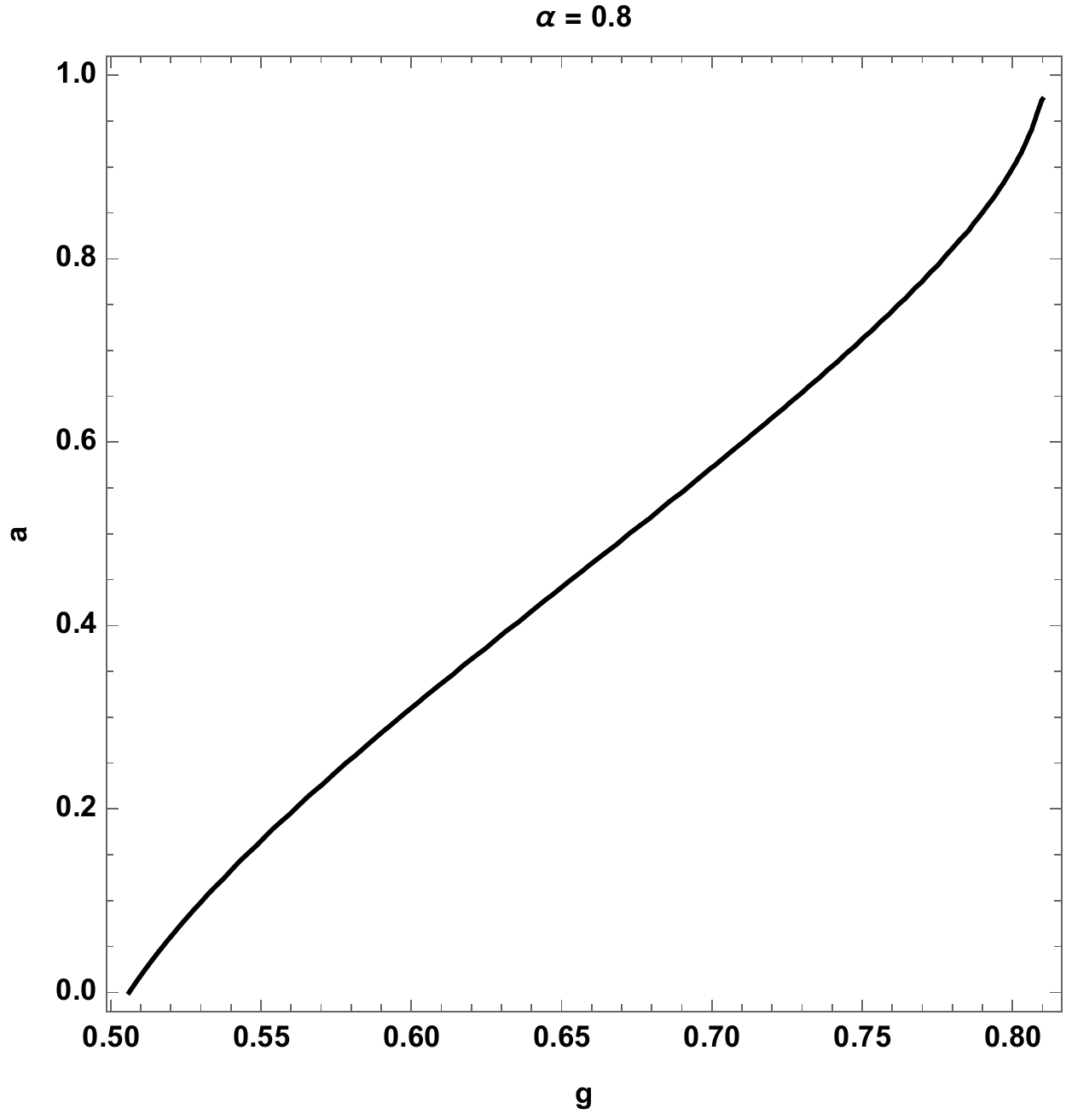}
	\end{center}
\caption{Relations among the rotation parameter $a$, the magnetic charge $g$, and the parameter $\alpha$ providing the same value of the ISCO radius.} \label{ag}
\end{figure*}

Based on these results, one may conclude that the magnetic charge parameter and the intensity of PFDM cannot completely mimic the rotation parameter of a Kerr black hole independently. However, one might expect that in some combinations of $g$ and $\alpha$ parameters, they can completely mimic the rotation parameter or at least up to the values very close to $1$. To see how well they can do degeneracy, we may plot the degeneracy between discussed parameters as plotted in Fig. ~\ref{ag}. In the left and central panels, it has been plotted how these parameters can mimic the rotation parameter when one of the two parameters, $g$ or $\alpha$ is absent. In the right one, it has been shown that for fixed value of the intensity of PFDM the magnetic charge of a black hole can almost completely mimic the spin parameter of a Kerr black hole. {It comes to the assumption that black holes in the Universe might be magnetically charged ones with dark matter surrounded rather than spinning Kerr ones and it would be difficult to distinguish them if one relies on the measurements that have a dependence on the ISCO radius for the test particles. Measurements of the radiative efficiency at the ISCO and Doppler boosting at the ISCO may be good examples.} 

\begin{figure*}[t!]
	\begin{center}
		a.
		\includegraphics[width=0.45\linewidth]{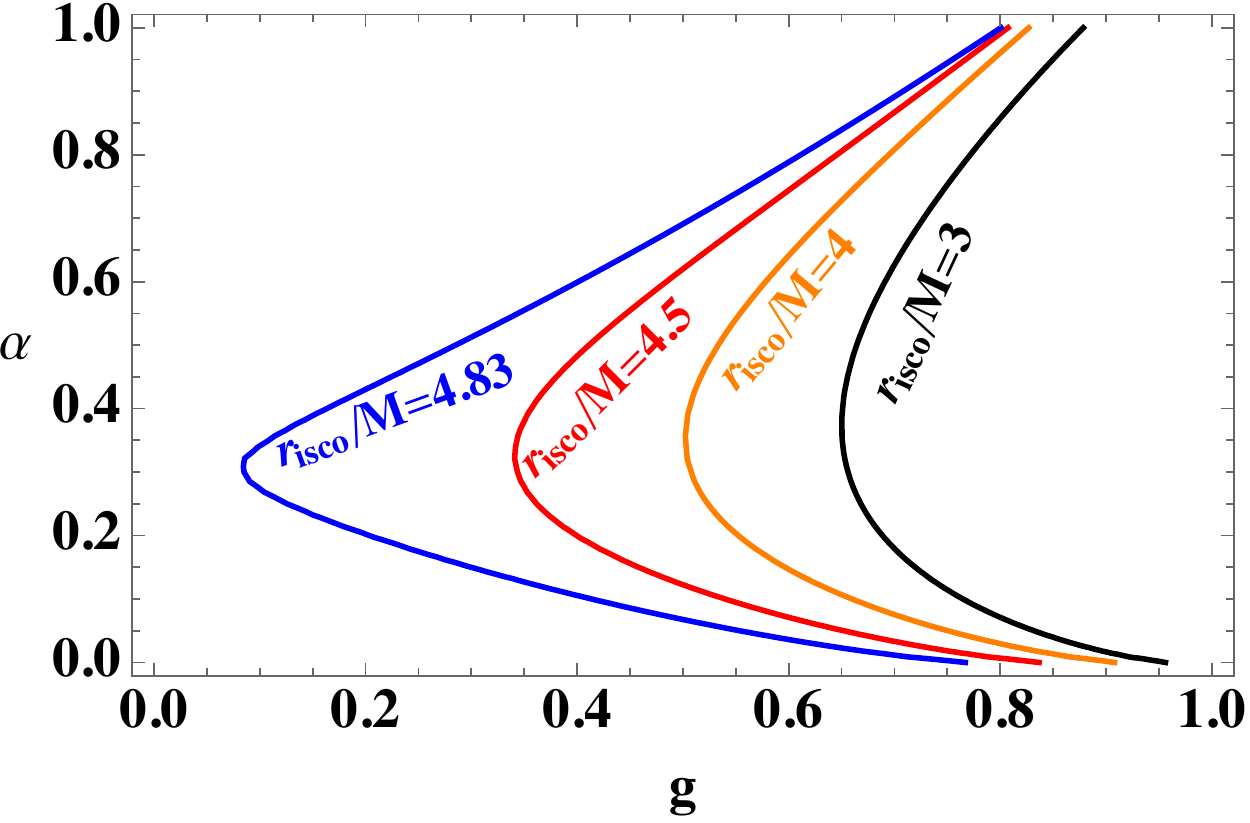}
		b.
		\includegraphics[width=0.45\linewidth]{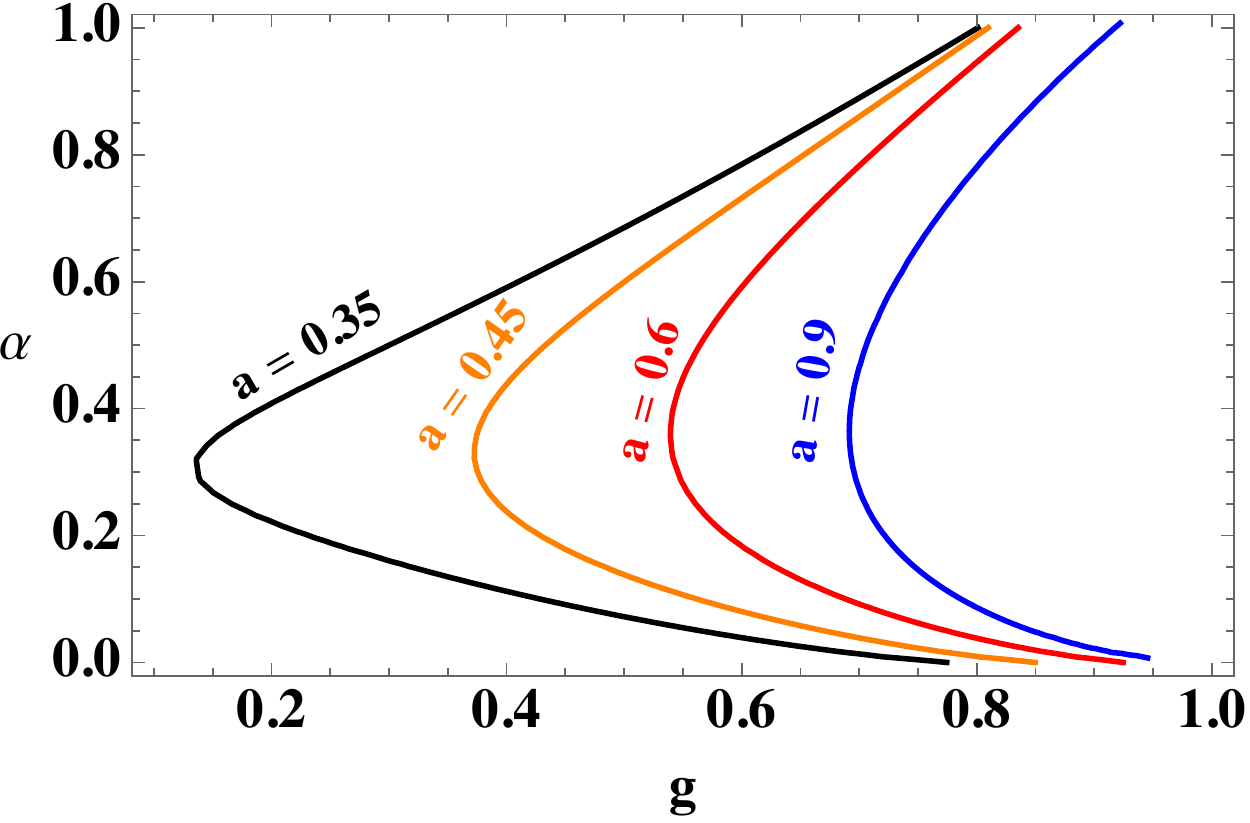}
	\end{center}
	\caption{The degeneracy plot between parameter $\alpha$ and magnetic charge $g$ for fixed chosen values of a. ISCO radii $r_{ISCO}$ and b. spin parameter $a$.} \label{aa_g}
\end{figure*}

One can also plot the degeneracy between two parameters of the spacetime metric, $g$ and $\alpha$, which is shown in Fig. ~\ref{aa_g}.

\section{\label{Sec:motion} Magnetically charged particle motion around Bardeen black hole surrounded by perfect fluid dark matter}

Now we study magnetically charged test particle with the rest mass $m$ and magnetic charge $q_m$ around Bardeen black hole surrounded by perfect fluid dark matter. We define the general form of Hamiltonian~\cite{Misner73}, which, respectively, describes a particle motion  of electrically and magnetically charged particles around Bardeen regular black hole as 
\begin{eqnarray}
 H  \equiv  \frac{1}{2}& g^{\alpha\beta}&\left(\frac{\partial {\cal S}}{\partial
x^{\alpha}}-q{A}_{\alpha}+iq_m{A}^{\star}_{\alpha}\right)\nonumber\\  &\times &\left(\frac{\partial {\cal S}}{\partial
x^{\beta}}-q{A}_{\beta}+iq_m{A}^{\star}_{\beta}\right)\, ,
\label{Eq:H}
\end{eqnarray}
where ${\cal S}$ is the action and $q$ and $q_m$ are the electric and magnetic charges of falling in test particle, respectively. Note that the components of the vector and the dual vector potentials $A_{\alpha}$ and $A^{\star}_{\alpha}$ of the given electromagnetic field read
\begin{eqnarray}
A^{\star}_t=-\frac{ig}{r}\,  \mbox{~~~and~~~} A_{\phi}=-g\cos\theta\,  .
\end{eqnarray}
Here we should note that the test particle has only a nonvanishing magnetic charge $q_m$, i.e. $q=0$, and hence we consider only the nonvanishing $A_t$ component of the electromagnetic field.

According to the well-known properties of Hamiltonian, we have $H=k/2$ in relation to $k=-m^2$, where $m$ is the mass of test particle with magnetic charge. Let us then write the action ${\cal S}$ for Hamilton-Jacobi equation describing the magnetically charged test particle motion around Bardeen regular black hole as 
\begin{eqnarray}\label{Eq:separation1}
{\cal S}= -\frac{1}{2}k\lambda-Et+L\varphi+{\cal S}_{r}(r)+{\cal S}_{\theta}(\theta)\ .
\end{eqnarray}
The Hamilton-Jacobi equation then yields  

\begin{eqnarray}\label{Eq:separable}
k&=& -
\frac{1}{f(r)}\left[-E+
\frac{q_m\,g}{r} \right]^{2}
+f(r)\,
\left(\frac{\partial S_{r}}{\partial
r}\right)^{2}\nonumber\\&+&{1\over r^{2}}\left(\frac{\partial S_{\theta}}{\partial
\theta}\right)^{2}+\frac{L^{2}}{r^{2} \sin^2\theta } \, . 
\end{eqnarray}
In the above equation we have $E$, $L$ and $k$ being independent constants of motion as well as the constant of the latitudinal motion. The fourth one is caused by the separability of the action. Since we restrict ourselves to the equatorial plane (i.e. $\theta=\pi/2$) we are further able to eliminate the fourth constant of motion~\cite{Misner73}.
\begin{figure*}
\centering
 \includegraphics[width=0.3\textwidth]{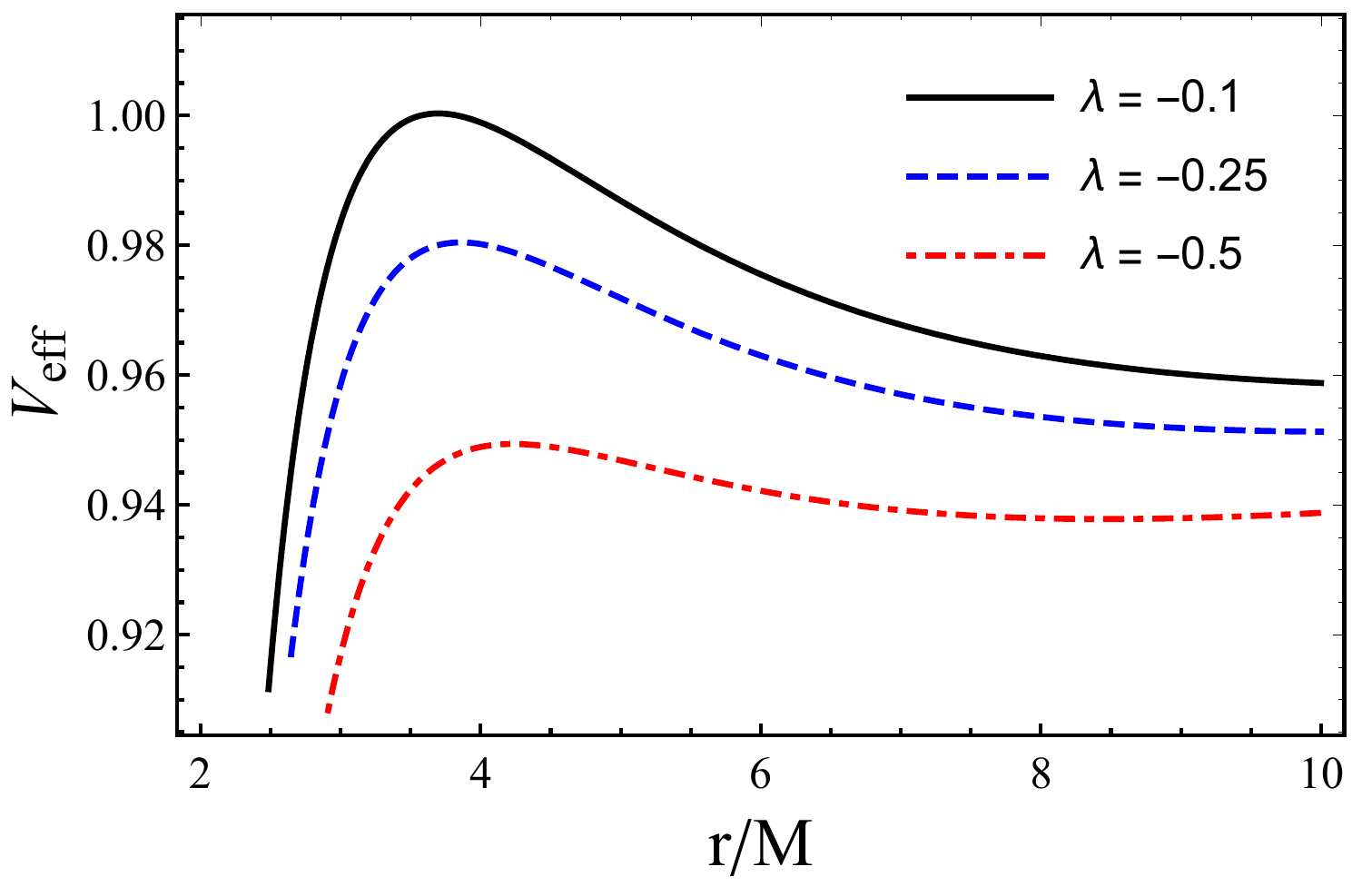}%
 \includegraphics[width=0.3\textwidth]{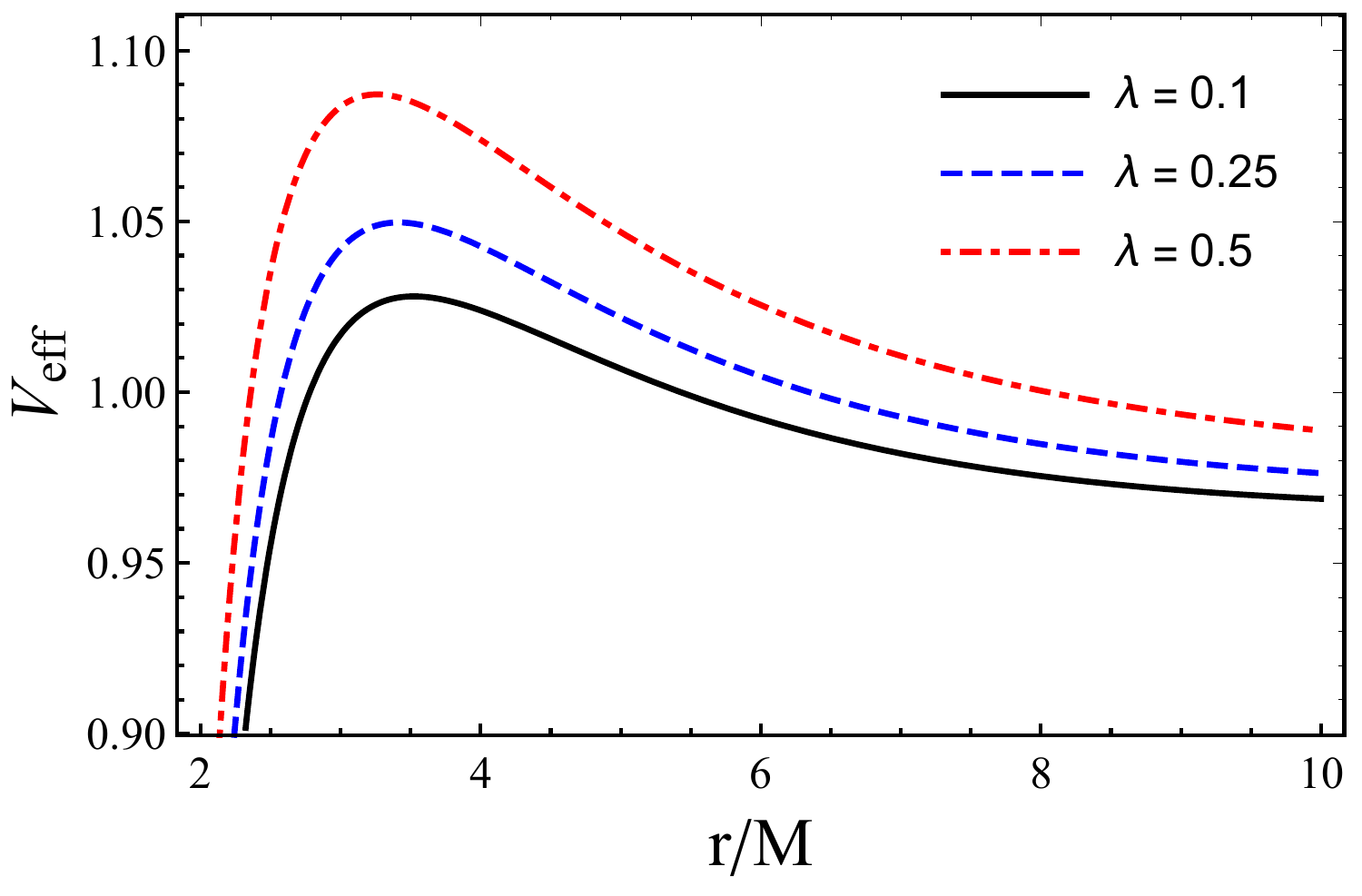}
 \includegraphics[width=0.3\textwidth]{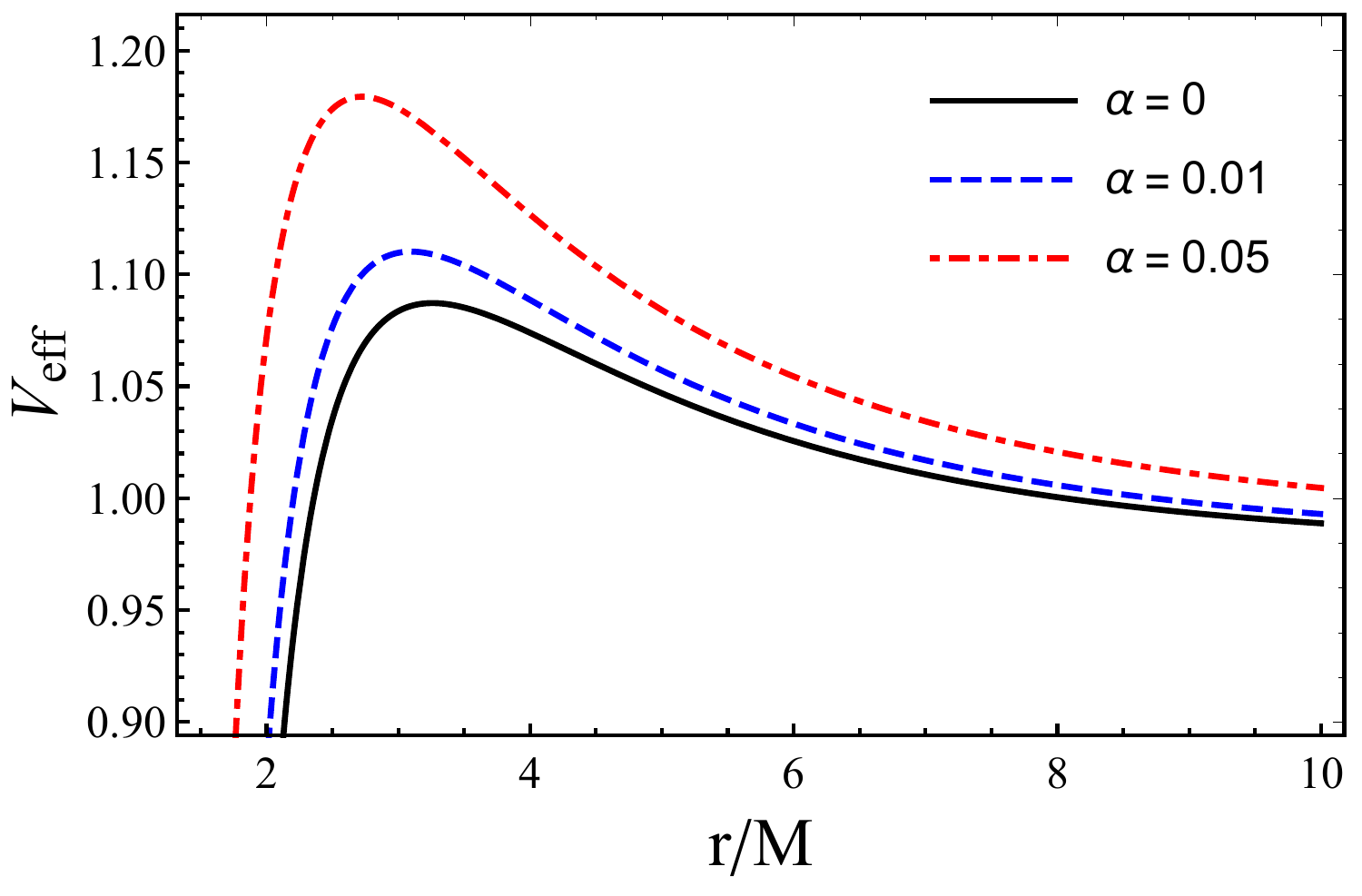}
\caption{\label{fig:eff} Radial dependence of the effective potential for massive magnetically charged particles around the Bardeen regular black hole with magnetic charge $g$. Left and middle panels: Veff against $r/M$ for different values of negative and positive magnetic charge parameter $\lambda$ in the case of fixed $g=0.5$ and $\alpha=0.0$. Right panel: For different values of perfect fluid dark matter parameter $\alpha$ in the case of fixed $g = 0.5$ and $\lambda= 0.5$.
}
\end{figure*}
\begin{figure*}

  \includegraphics[width=0.3\textwidth]{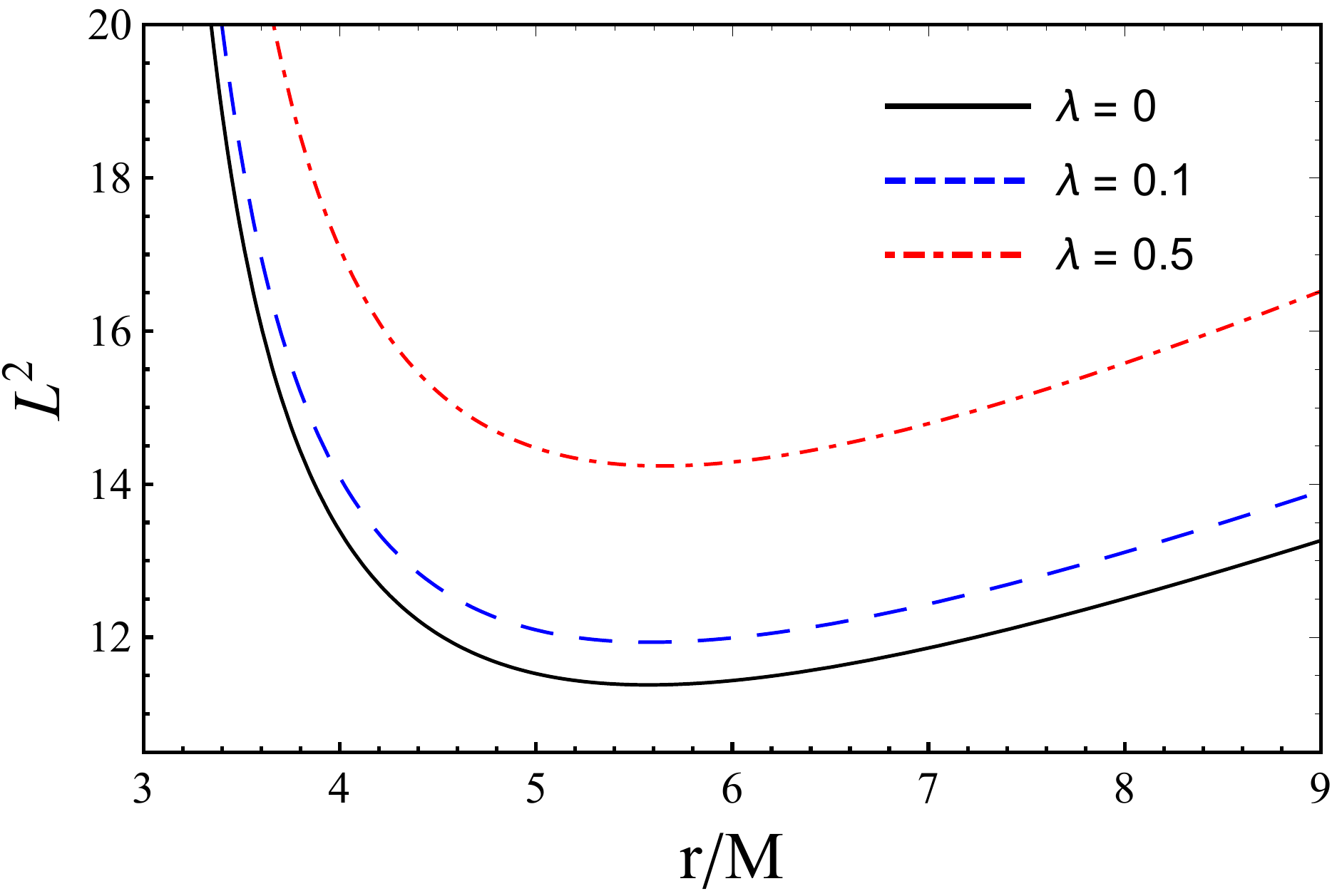}%
  \includegraphics[width=0.3\textwidth]{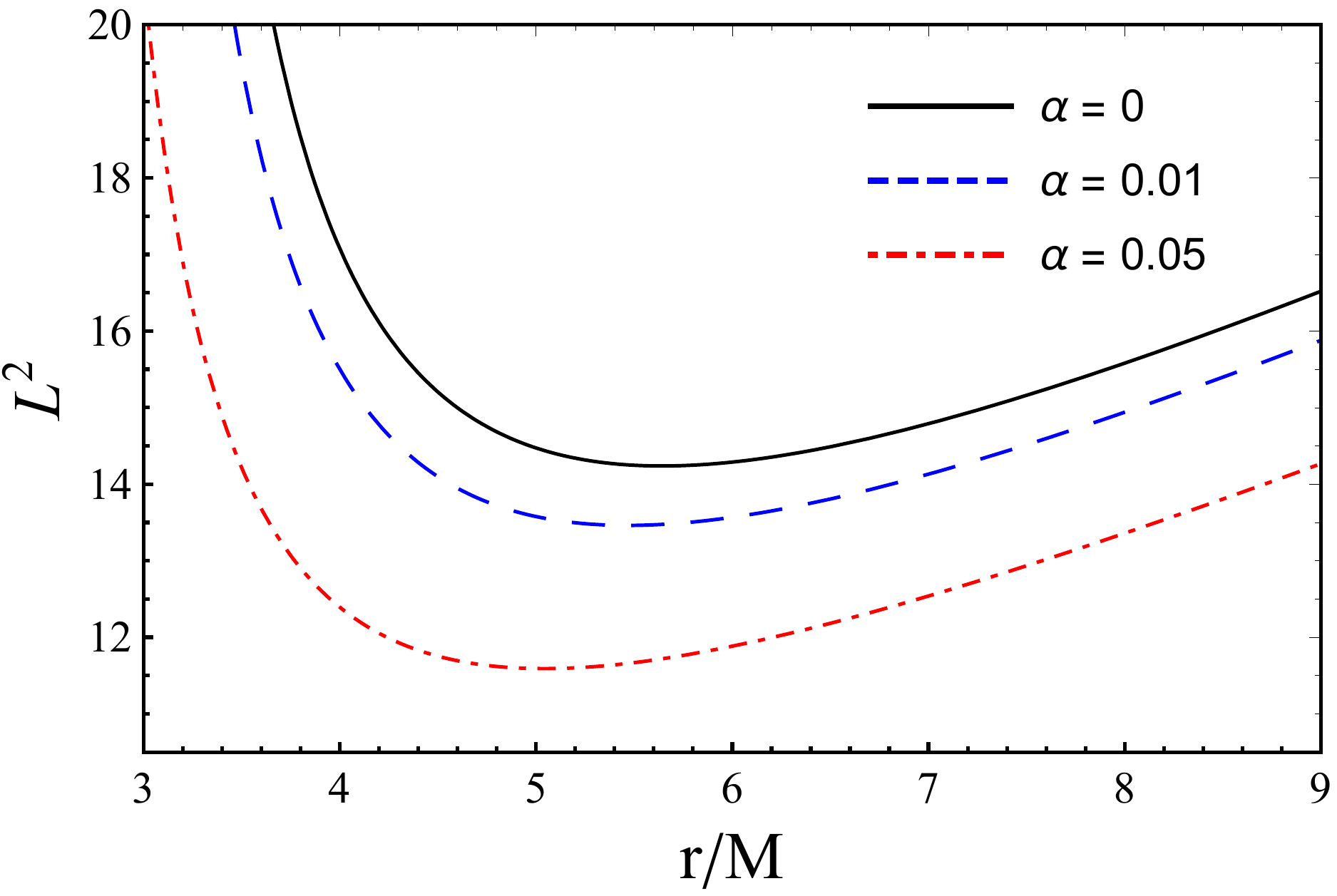}
 \includegraphics[width=0.3\textwidth]{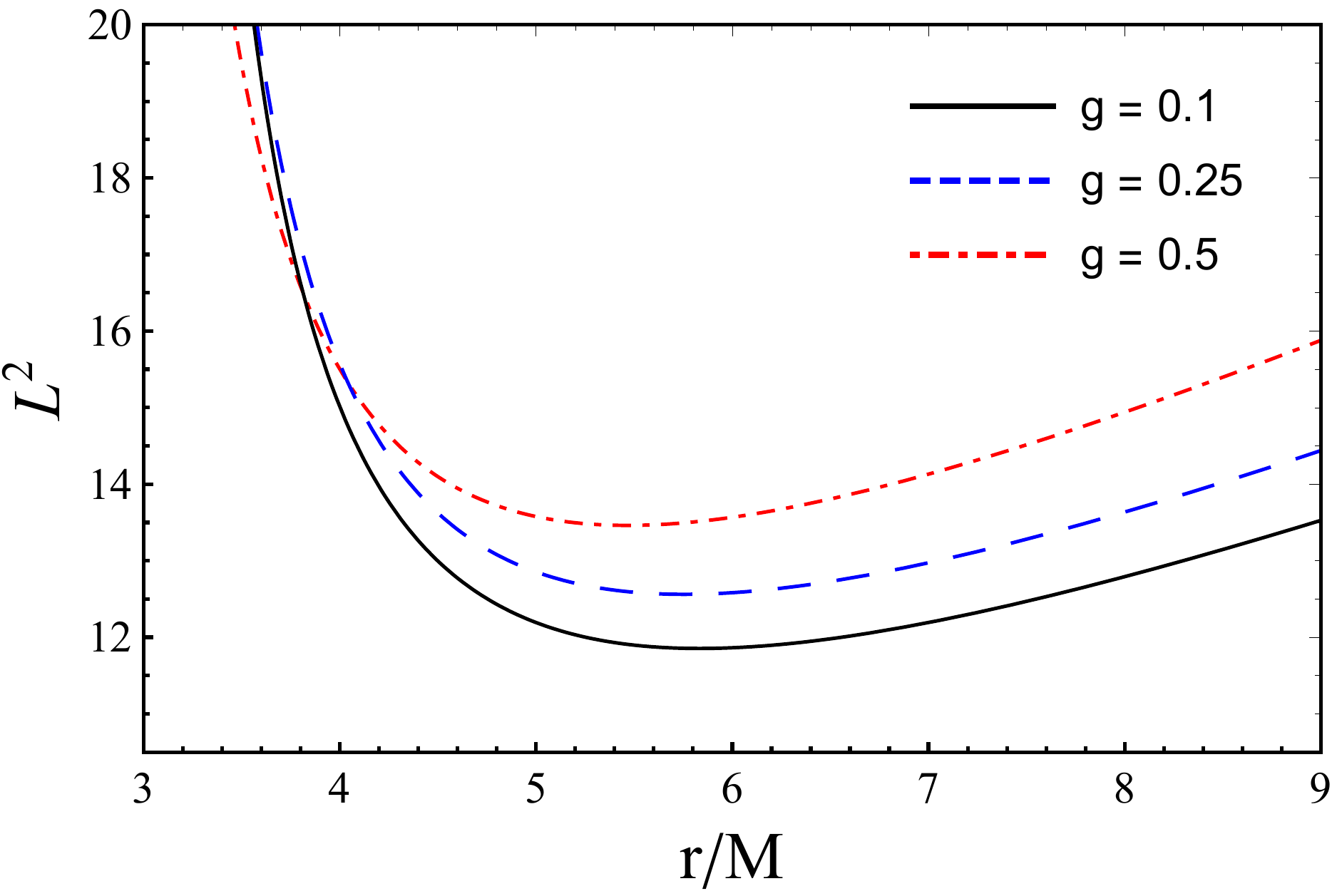}
\caption{\label{fig:ang} The radial dependence of the specific angular momentum for magnetically charged test particles around the Bardeen regular black hole with magnetic charge $g$. Left panel:  For the different values of magnetically charge parameter $\lambda$ in the case of fixed $g=0.5$ and $\alpha=0$. Right panel: For for different values of perfect fluid dark matter parameter $\alpha$ in the case of fixed $g=0.5$ and $\lambda=0.5$. }
\end{figure*}
\begin{table}[h]
\caption{\label{tab1} The value of the ISCO radius of the magnetically charged particles for the
different values of magnetic charge parameter $\lambda$ and perfect fluid dark matter parameter $\alpha$ for fixed values of $g=0.7$.  }
\begin{ruledtabular}
\begin{tabular}{c|ccc}
$ $   & $ $ &  $\lambda$
& $ $   \\ \hline
{$\rm \alpha$}  & $0.1$ &  $0.25$
& $0.50$   \\
         \hline\\
0.005              &4.91080  &4.80798  &4.79484 \\\\
0.010            &4.80708  &4.70621  &4.68143 \\\\
0.050          &4.13641  &4.04891  &3.89389  \\
\end{tabular}
\end{ruledtabular}
\end{table}
\begin{figure*}

  \includegraphics[width=0.45\textwidth]{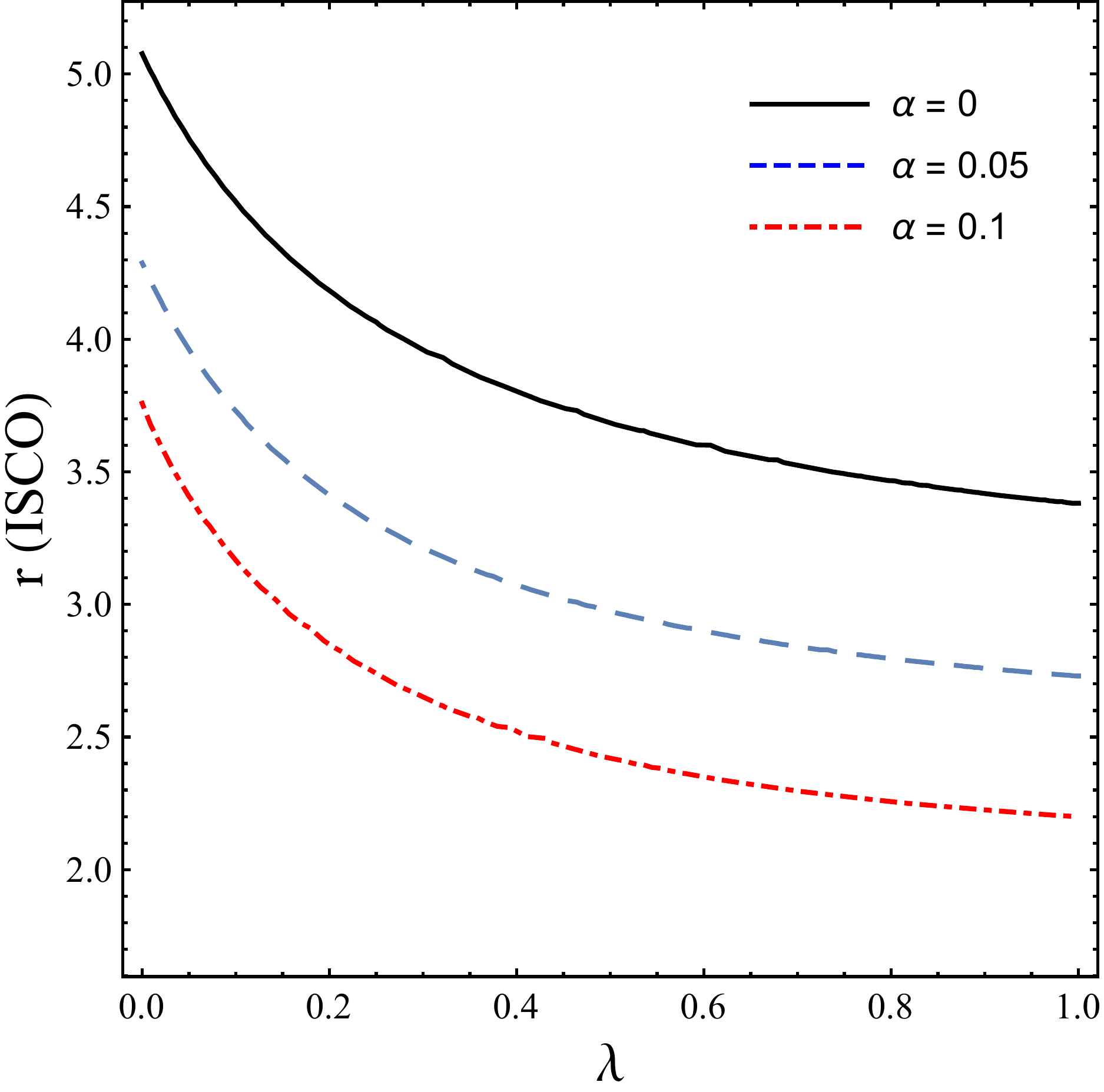}
  \includegraphics[width=0.45\textwidth]{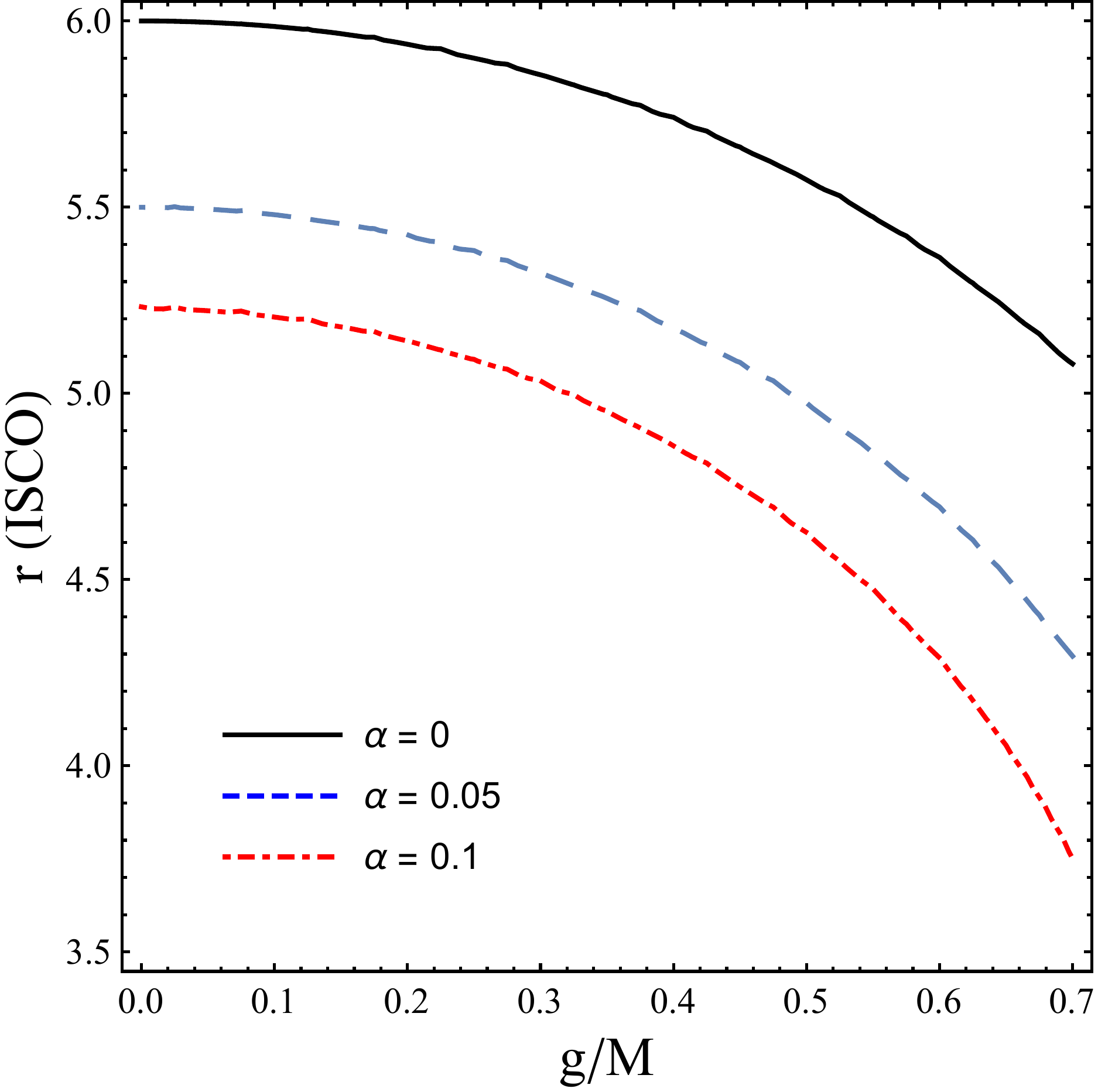}

\caption{\label{fig:isco} The dependence of the ISCO radius from the parameters $\lambda$ and $g$. Left panel: For the different values of perfect fluid dark matter parameter $\alpha$ in the case of fixed $g=0.7$. Right panel: For $\lambda=0$ for the different values of parameter $\alpha$. }
\end{figure*}
From Eq.~(\ref{Eq:separable}), it is then straightforward to obtain the radial equation of motion for magnetically charged particles in the following form
\begin{eqnarray}\label{Eq:rdot}
\dot{r}^2=\Big(\mathcal{E}-\mathcal{E}_-(r)\Big)\Big(\mathcal{E}-\mathcal{E}_+(r)\Big)\, ,
\end{eqnarray}
with the radial function $\mathcal{E}_{\pm}(r,\mathcal{L},g,\alpha,\lambda)$
of the radial motion  
\begin{eqnarray} \label{Veff2}
\mathcal{E}_{\pm}(r,\mathcal{L},g,\alpha,\lambda)&=& \frac{\lambda g}{r} \pm
\left(1+\frac{\mathcal{L}}{r^{2}}\right)^{1/2}\nonumber\\ &\times &\left(1-\frac{2\, M r^2}{\left(r^2+g^2\right)^{3/2}}+\frac{\alpha}{r}\ln\frac{r}{\vert\alpha \vert}\right)^{1/2}\, , \nonumber\\
\end{eqnarray}
with $\mathcal{E}=E/m$,
$\mathcal{L}=L/m$, $\lambda=q_m/m$ and $k/m^2=-1$. From Eq.~(\ref{Eq:rdot}), $\dot{r}^2\geq 0$ always, thus leading to either $\mathcal{E}>\mathcal{E}_{+}(r,\mathcal{L},g,\alpha,\lambda)$ or $\mathcal{E}<\mathcal{E}_{-}(r,\mathcal{L},g,\alpha,\lambda)$. However, we shall focus on $\mathcal{E}_{+}(r,\mathcal{L},g,\alpha,\lambda)$ which gives the turning points of the radial motion and its properties and consequently we select it as a effective potential (i.e. $V_{\rm eff}(r,\mathcal{L},g,\alpha,\lambda)$). In the limiting case eliminating perfect fluid dark matter parameter $\alpha$ and magnetic charge $g$ of the black we obtain the radial motion of the test particle for the Schwarzschild black hole. 

Now we deal with the effective potential $V_{\rm eff}(r,\mathcal{L},g,\alpha,\lambda)$ for the radial motion of magnetically charged test particles. Fig.~\ref{fig:eff} reflects the role of both negative and positive magnetic charge parameter $\lambda$ for the radial profiles of the effective potential for fixed black hole's magnetic charge parameter $g$ in the case of vanishing perfect fluid dark matter parameter $\alpha$, while the last panel reflects the role of the perfect fluid dark matter for fixed values of $g$ and $\lambda$.  It is easily seen from the radial dependence of  $V_{\rm eff}(r,\mathcal{L},g,\alpha,\lambda)$, the height and range of the effective potential decrease with increasing the value of negative magnetic charge parameter $\lambda<0$ while this behavior is opposite with increasing the value of the positive $\lambda>0$ in the case of fixed $g$ and vanishing perfect fluid dark matter parameter $\alpha$.  In a similar way, we observe that the range of the effective potential increases with increasing perfect fluid dark matter parameter $\alpha$ while the other parameters are fixed. It is worth mentioning that the strength of the potential for positive $\lambda$ becomes stronger in comparison with one for negative values of parameter $\lambda$ from Fig.~\ref{fig:eff}. 

Now, we come to the study of the circular orbits of magnetically charged test particles around Bardeen regular black hole having magnetic charge.  We explore required conditions to understand whether test particles are able to be on the circular orbits or not. Thus, we have
\begin{eqnarray}\label{Eq:cir1}
V_{\rm eff}(r,\mathcal{L},g,\alpha,\lambda)=\mathcal{E}\ , \\ \nonumber\\ V_{\rm eff}^{\prime}(r,\mathcal{L},g,\alpha,\lambda)=0\, ,
\label{Eq:cir2}
\end{eqnarray}
where prime denotes a derivative with respect to $r$. As always, we then determine specific energy $\mathcal{E}$ and angular momentum $\mathcal{L}$ for magnetically charged particles moving at the circular orbits,
%
\begin{eqnarray}
\mathcal{E}&=&\frac{\lambda g}{r} +
\sqrt{f(r)\left(1+\frac{\mathcal{L}^2}{r^{2}}\right)}\, , \\
\mathcal{L}^2&=& \frac{r^3 f'(r) \Big(2 f(r)-r f'(r)\Big)+2 g^2 \lambda ^2 f(r)}{\Big(r f'(r)-2 f(r)\Big)^2}\nonumber\\&+& \frac{g \lambda  f(r) \sqrt{g^2 \lambda ^2+4 r^2 f(r)-2 r^3 f'(r)}}{\Big(r f'(r)-2 f(r)\Big)^2}
 \, .
\end{eqnarray}
%
The radial profiles of the specific angular momentum of the magnetically charged particle for various cases is illustrated in Fig.~\ref{fig:ang} for circular orbits around the black hole. We can see from Fig.~\ref{fig:ang} that  circular orbit of the magnetically charged particle shifts outward the central object, i.e. toward the large radii with increasing magnetic charge $\lambda$ for fixed $g$ and vanishing perfect fluid dark matter parameter $\alpha$. However, this manner is compensated by the perfect fluid dark matter. Also, we can observe that the circular orbits become close to the black hole horizon with increase in the value of black hole charge parameter $g$.

We shall now consider in particular study the ISCO for magnetically charged test particles moving in the black hole spacetime. For determining the ISCO, the strong condition is required, and hence we solve the equation for second derivative of the effective potential
\begin{eqnarray}\label{Eq:is}
V_{\rm eff}^{\prime\prime}(r,\mathcal{L},g,\alpha,\lambda)=0\, .
\end{eqnarray}
In Table~\ref{tab1}, we provide the ISCO radius obtained by solving Eqs.~(\ref{Eq:cir2}) and (\ref{Eq:is}) numerically for different values of perfect fluid dark matter parameter $\alpha$ and magnetic charge parameter $\lambda$ in the case of fixed black hole magnetic charge parameter $g$. The behavior of the results tabulated in Table~\ref{tab1} is also reflected in Fig.~\ref{fig:isco}. As can be seen from Fig.~\ref{fig:isco}, the ISCO radius decreases due to the presence of perfect fluid dark matter.      

In the case of neutral test particle $\lambda=0$, we consider the ISCO radius for which the minimum value of the angular momentum from $V_{eff}^{\prime}(r,\mathcal{L},g,\alpha,\lambda)=0$ reads 
\begin{eqnarray}\label{Eq:min-ang}
\mathcal{L}^2=\frac{r^3 f'(r)}{2 f(r)-r f'(r)}\, ,
\end{eqnarray}
\\ 
for which particles are allowed to be on circular orbits. From the condition $V_{\rm eff}^{\prime\prime}(r,\mathcal{L},g,\alpha,\lambda)=0$ we obtain $r_{ISCO}$ by virtue of Eq.~(\ref{Eq:min-ang}).  
In Fig.~\ref{fig:isco}, on the right panel, we show the relation between ISCO radius and black hole magnetic charge parameter $g$ for the different values of perfect fluid dark matter parameter $\alpha$. It is easily seen that, in the case of $\lambda=0$, $r_{ISCO}$ decreases with increasing $g$ and $\alpha$ as well. Thus, the ISCO is strongly influenced due to the combined effect of perfect fluid dark matter and black hole magnetic charge parameter. 

\begin{figure}

 \includegraphics[width=0.45\textwidth]{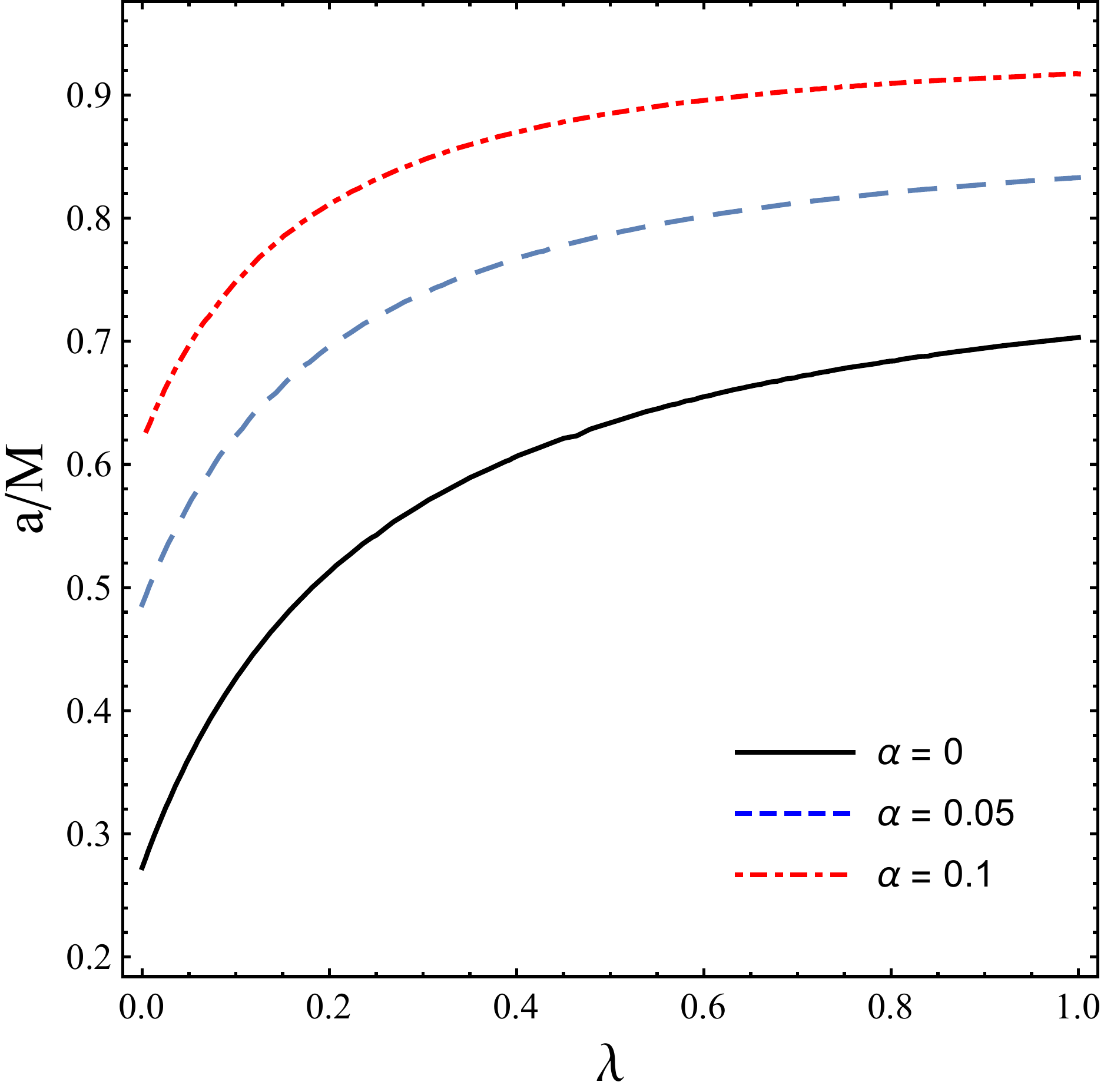}

\caption{\label{fig:a_vs_g} 
The plot illustrates the values of rotation parameter $a$ as a function of the magnetic charge parameter $\lambda$ for which the ISCO radius behaves the same in both the Kerr geometry and the Bardeen black hole geometry surrounded by perfect fluid dark matter for different values of $\alpha$ for a given value of $g$. }
\end{figure}

In spite of the fact that there are the first observational facts of black hole existence by the LIGO and Virgo scientific collaborations \cite{LIGO16a,LIGO16} and the first image of supermassive black hole candidate, i.e. of the central object of the elliptical galaxy M87~\cite{EHT19a,EHT19b}, black holes have been still regarded as candidates due to the fact that their parameters except the total mass $M$ have not been explicitly detected yet. Thus, one can make an effort to predict black hole parameters using astronomical observations in black hole close environment. In this respect the presence of perfect fluid dark matter due to its attractive  effect and black hole magnetic charge would play a decisive role to understand the geodesics of test particles around black holes and black hole's angular momentum.  We now devote our attention to possibility to extract  the value of magnetic charge parameter and perfect fluid dark matter parameter from astronomical observations. For distant  observer it is not possible to distinguish between two geometries by analyzing electromagnet radiations emitted by accretion disk around black hole.  From Fig.~\ref{fig:a_vs_g} one can observe that combined effect of perfect fluid dark matter and magnetic charge parameter can mimic black hole rotation parameter up to $a/M\approx 0.9$, whereas in the case of vanishing $\alpha=0$ black hole magnetic charge alone can mimic only up to  $a/M\approx 0.6-0.65$.  However, in spite that is an idealized model, yet it helps to understand a whole process around Bardeen regular black hole with magnetic charge. 

\section{magnetic dipole motion  \label{section3}}

 The orthonormal radial component of magnetic field generated by the magnetic charge of Bardeen regular black hole is 
\begin{equation}\label{BrBt}
    B^{\hat{r}}=\frac{g}{r^2} \ .
    \end{equation}
 The radial component of the magnetic field around magnetically charged black hole formally coincides with the standard Newtonian expression.
    Now, one may study the dynamics of magnetic dipoles around magnetically charged black holes using the Hamilton-Jacobi equation \cite{deFelice}
\begin{eqnarray}\label{HJ}
g^{\mu \nu}\frac{\partial {\cal S}}{\partial x^{\mu}} \frac{\partial {\cal S}}{\partial x^{\nu}}=-\Bigg(m-\frac{1}{2} {\cal D}^{\mu \nu}F_{\mu \nu}\Bigg)^2\ ,
\end{eqnarray}
with the term ${\cal D}^{\mu \nu}{\cal F}_{\mu \nu}$ being responsible for the interaction between the magnetic dipoles and the magnetic field generated by magnetic charge of the Bardeen regular black hole.


Here we assume that the magnetic dipole is described by the corresponding polarization tensor ${\cal D}^{\alpha \beta}$ that satisfies the following condition
\begin{eqnarray}\label{dexp}
{\cal D}^{\alpha \beta}=\eta^{\alpha \beta \sigma \nu}u_{\sigma}\mu_{\nu}\ , \qquad {\cal D}^{\alpha \beta }u_{\beta}=0\ ,
\end{eqnarray} 
where $\mu^{\nu}$ is dipole moment of the magnetic dipole. Here we determine the interaction term ${\cal D}^{\mu \nu}{\cal F}_{\mu \nu}$ using the relation between the electromagnetic field tensor $F_{\alpha \beta}$ and components of electric $E_{\alpha}$ and magnetic $B^{\alpha}$ fields as 
\begin{eqnarray}\label{fexp}
F_{\alpha \beta}=w_{\alpha}E_{\beta}-w_{\beta}E_{\alpha}-\eta_{\alpha \beta \sigma \gamma}w^{\sigma}B^{\gamma}\ ,
\end{eqnarray}
where $w_{\alpha}$ is the four-velocity of the observer measuring the electromagnetic field. Taking into account the condition given in (\ref{dexp}) and non-zero components of the electromagnetic field tensor we have 
\begin{eqnarray}\label{DF1}
{\cal D}^{\alpha \beta}F_{\alpha \beta}=2\mu_{\alpha}B^{\alpha}=2{\cal \mu}^{\hat{\alpha}}B_{\hat{\alpha}} \ .
\end{eqnarray}

We assume that the direction of the magnetic moment of the magnetic dipole lies on the equatorial plane being parallel to the magnetic field generated by the regular Bardeen black hole. It is indeed favorable direction for the magnetic moment and consistent with the lowest energy level for the magnetic dipole. In turn, it leads the magnetic moment to have orthonormal components as $\mu^{\hat{i}}=(\mu^{\hat{r}},0,0)$. Consequently, taking into account Eqs. (\ref{DF1}) and (\ref{BrBt}) one may rewrite the interaction term in the following form
\begin{eqnarray}\label{DF3} {\cal D}^{\alpha \beta}F_{\alpha \beta} = \frac{2\mu g}{r^2} \ , \end{eqnarray}
where $\mu = \left(\vline \mu_{\hat{i}}\mu^{\hat{i}} \vline\right)^{1/2}$ is the norm of the magnetic dipole moment of the magnetic dipole.

Since, the spacetime symmetries are not changed by the existence of the axial symmetric proper magnetic field of regular Bardeen black hole, therefore we have two conserved quantities of motion of the magnetic dipole as energy $p_t = -E$ and angular momentum $p_{\phi}= L$. 
The radial motion of a magnetized particle around magnetically charged Bardeen regular black hole at the equatorial plane, where $\theta=\pi/2$, with $p_{\theta}=0$, using (\ref{DF1}), (\ref{HJ}) and the action (\ref{action1}) gives the following form 
\begin{eqnarray}
\dot{r}^2={\cal{E}}^2-V_{\rm eff}(r;l,{\cal B},\alpha)\ .
\end{eqnarray}
The effective potential has the form
\begin{eqnarray}\label{effpot}\nonumber
V_{\rm eff}(r;l,{\cal B},\alpha)&=&\left(1-\frac{2M r^2}{\left(r^2+g^2\right)^\frac{3}{2}}+\frac{\alpha}{r}\ln \frac{r}{|\alpha|} \right)\\
&\times &\left[\left(1-\frac{{\cal B}}{r^2}\right)^2+\frac{{\cal L}^2}{r^2}\right]\ ,
\end{eqnarray}
where the relation $${\cal B} =\frac{\mu}{m}g$$ is a new introduced parameter which is responsible for the interaction between dipole moment of magnetic dipole and the proper magnetic field of the magnetically charged regular Bardeen black hole in PFDM. We introduce a new parameter  $\beta=\mu/(m M)$ which illustrates parameters of the magnetic dipole and the central object and it takes only positive values for the case when neutron star is treated as a test magnetized particle orbiting a SMBH
\begin{eqnarray}
   \beta&=&\frac{B_{\rm NS}R_{\rm NS}^3}{2m_{\rm NS}M_{\rm SMBH}}\\\nonumber
   & \simeq & 0.18 \left(\frac{B_{\rm NS}}{10^{12}\rm G}\right)\left(\frac{R_{\rm NS}}{10^6 \rm cm}\right)\left(\frac{m_{\rm NS}}{M_{\odot}}\right)^{-1}\left(\frac{M_{\rm SMBH}}{10^6M_{\odot}}\right)^{-1}\ . 
\end{eqnarray}

For the system of the highly magnetized magnetar SGR (PSR) J1745--2900 with magnetic dipole moment $\mu \simeq 1.6\times 10^{32} \rm G\cdot cm^3$ and mass $m \approx 1.5 M_{\odot}$ orbiting around the SMBH Sagittarius A* (Sgr A*) ( $M \simeq 3.8 \times 10^6M_{\odot}$) ~\cite{Mori2013ApJ}, the value of the parameter $\beta$ can be easily estimated based on the observational data  as 
\begin{eqnarray}
 \beta=\frac{\mu_{\rm PSR\, J1745-2900}}{m_{\rm PSR\, J1745-2900} M_{\rm SgrA *}}\approx 10.2\ .
 \end{eqnarray} 
 
The circular stable orbits of the magnetic dipole around the central magnetized object can be defined by the standard conditions as
\begin{eqnarray} \label{conditions}
V_{\rm eff}'=0\ , V_{\rm eff}'' \geq 0  \ .
\end{eqnarray}
Specific angular momentum and energy of the magnetic dipole responsible for circular orbits can be expressed by the following expressions
\begin{widetext}
\begin{eqnarray}
\nonumber
{\cal L}^2&=&\frac{r^2 \left(1-\frac{\beta  g M}{r^2}\right)}{2-\frac{\alpha}{r}\left(1-3 \ln \frac{r}{\left| \alpha \right|}\right)-\frac{6 M r^4}{\left(g^2+r^2\right)^{5/2}}} \\
&\times & \Bigg\{ \left(1-\frac{\beta  g M}{r^2}\right) \left[\frac{\alpha}{r}  \left(1-\ln \frac{r}{\left| \alpha \right|}\right)+\frac{2 M r^2 \left(r^2-2 g^2\right)}{\left(g^2+r^2\right)^{5/2}}\right]+\frac{4 \beta  g M}{r^2} \left[1-\frac{2 M r^2}{\left(g^2+r^2\right)^{3/2}}+\frac{\alpha}{r}\ln\frac{r}{\left| \alpha \right| }\right] \Bigg\} \ ,
\end{eqnarray}
\begin{eqnarray}
\nonumber
{\cal E}^2&=&\left(1-\frac{\beta  g M}{r^2}\right) \left[1-\frac{2 M r^2}{\left(g^2+r^2\right)^{3/2}}+\frac{\alpha}{r}  \ln\frac{r}{\left| \alpha \right|}\right] \\
&\times &\Bigg\{ 1-\frac{\beta  g M}{r^2}+ \frac{\left(1-\frac{\beta  g M}{r^2}\right) \left[\frac{\alpha}{r}  \left(1-\ln\frac{r}{\left| \alpha \right| }\right)+\frac{2 M r^2 \left(r^2-2 g^2\right)}{\left(g^2+r^2\right)^{5/2}}\right]+\frac{4 \beta  g M}{r^2}\left[1-\frac{2 M r^2}{\left(g^2+r^2\right)^{3/2}}+\frac{\alpha}{r} \ln\frac{r}{\left| \alpha \right| }\right]}{2-\frac{\alpha}{r}  \left(1-3 \ln\frac{r}{\left| \alpha \right| }\right)-\frac{6 M r^4}{\left(g^2+r^2\right)^{5/2}}}\Bigg\} 
\ . 
\end{eqnarray}
\end{widetext}

\begin{figure}[h!]\centering\includegraphics[width=0.958\linewidth]{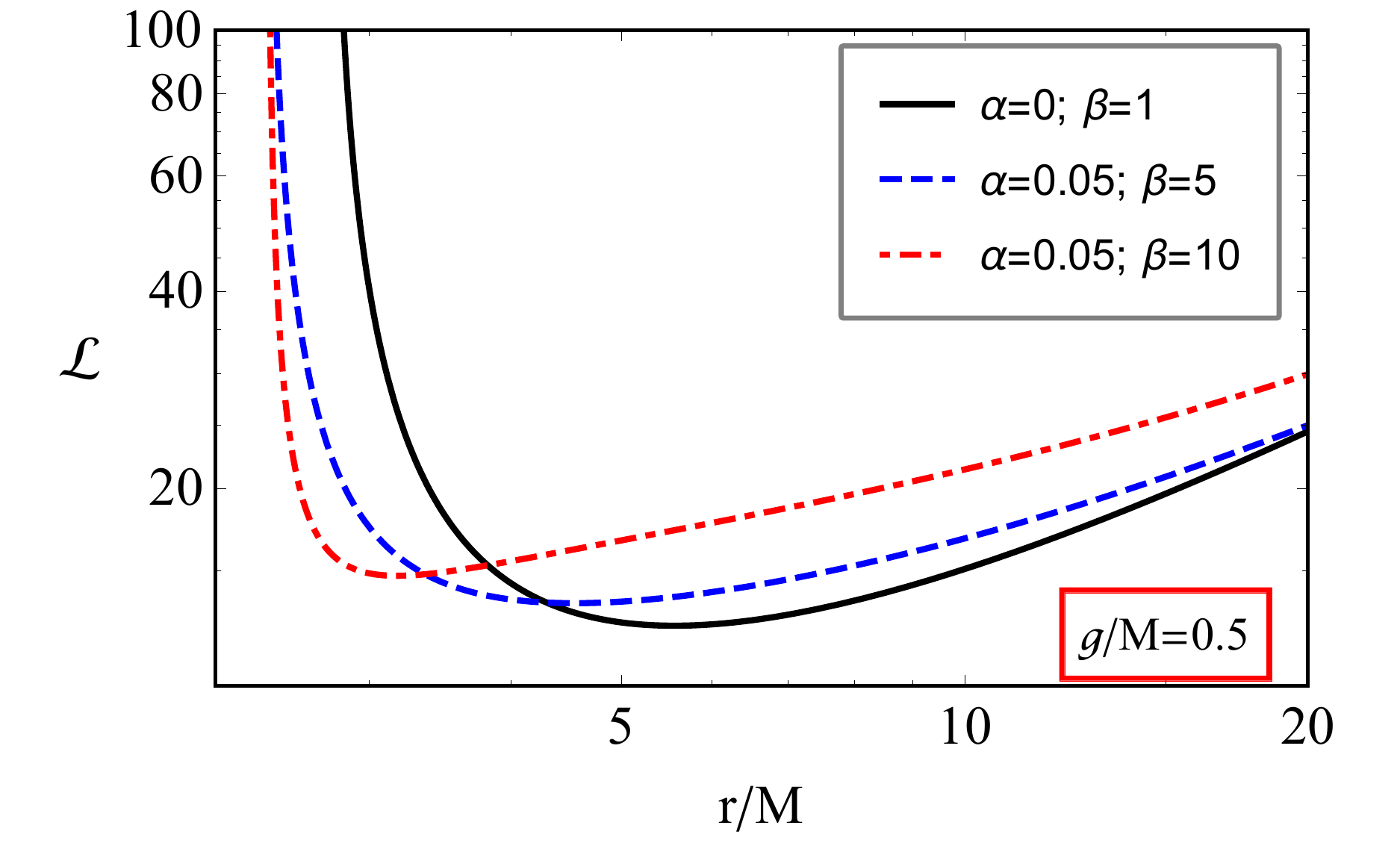}
\includegraphics[width=0.958\linewidth]{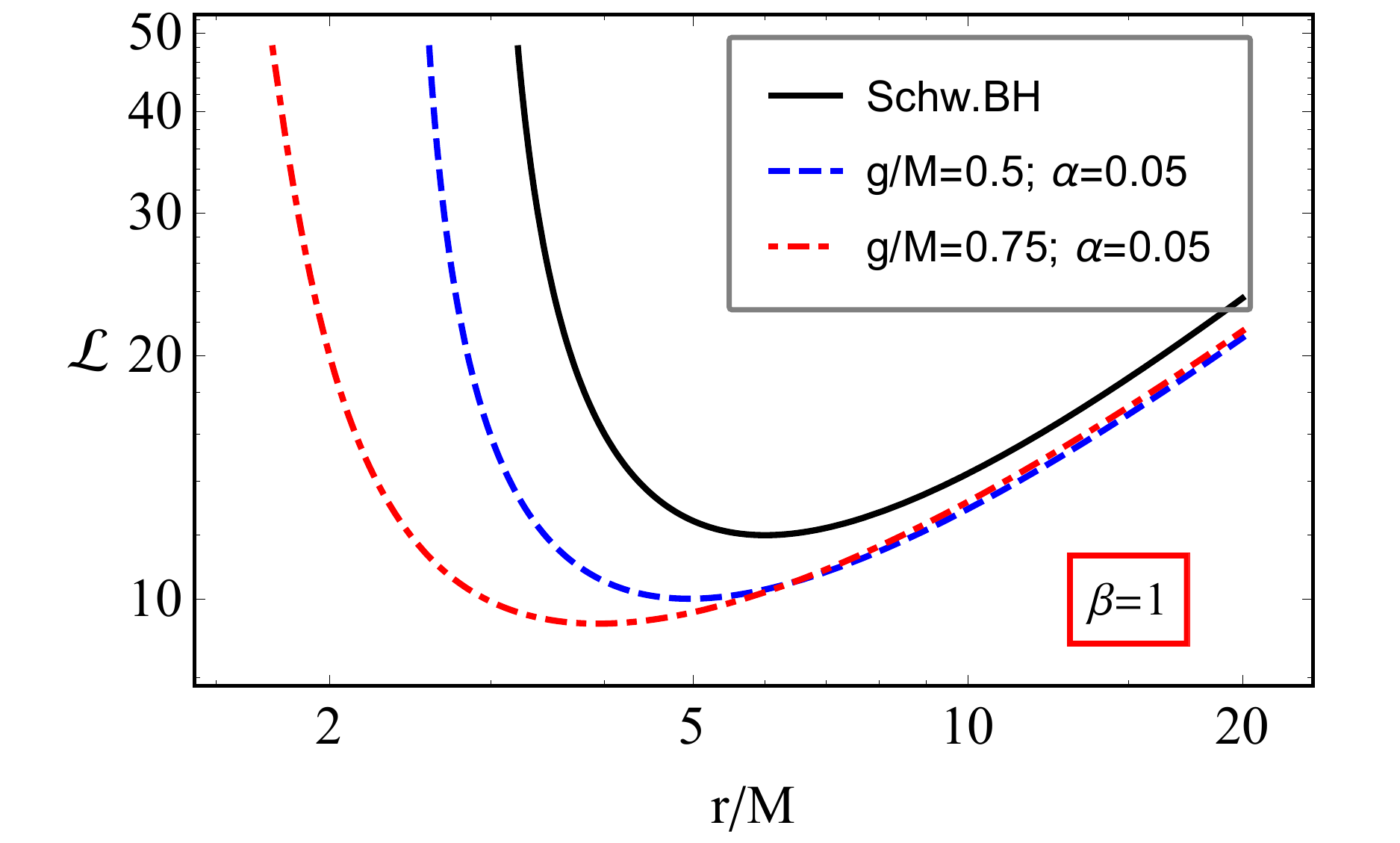} \caption{Specific angular momentum of the magnetic dipole for circular orbits as a function of radial coordinates for the different values of magnetic charge parameter $g$ and the parameter $\beta$. On the top panel, magnetic charge of the BH is fixed as $g=0.5M$ and in the bottom panel the parameter $\beta=1$. \label{LLbardeen}}\end{figure}

Radial dependence of the specific angular momentum of a magnetic dipole around the magnetically charged regular Bardeen black hole is demonstrated in Fig.\ref{LLbardeen}. On the top panel of the figure, the value of the magnetic charge parameter of the BH is fixed as $g/M=0.5$ (taken as $g<g_{\rm cr}\simeq0.78$) for the different values of the parameter of dark matter fluid $\alpha$ and on the bottom one we have fixed the parameter  $\beta$ as $\beta=1$ with the comparison to the  Schwarzschild case. One can see that the minimum value of the specific angular momentum increases and the distance where the angular momentum takes the minimum decreases with the increase of both the parameter of the dark matter fluid and the parameter $\beta$. Moreover, the increase of the magnetic charge parameter of the black hole causes to decrease the minimum values of the specific angular momentum and the distance where the angular momentum takes the minimum which corresponds to ISCO radius.

\begin{figure}[h!]\centering\includegraphics[width=0.9\linewidth]{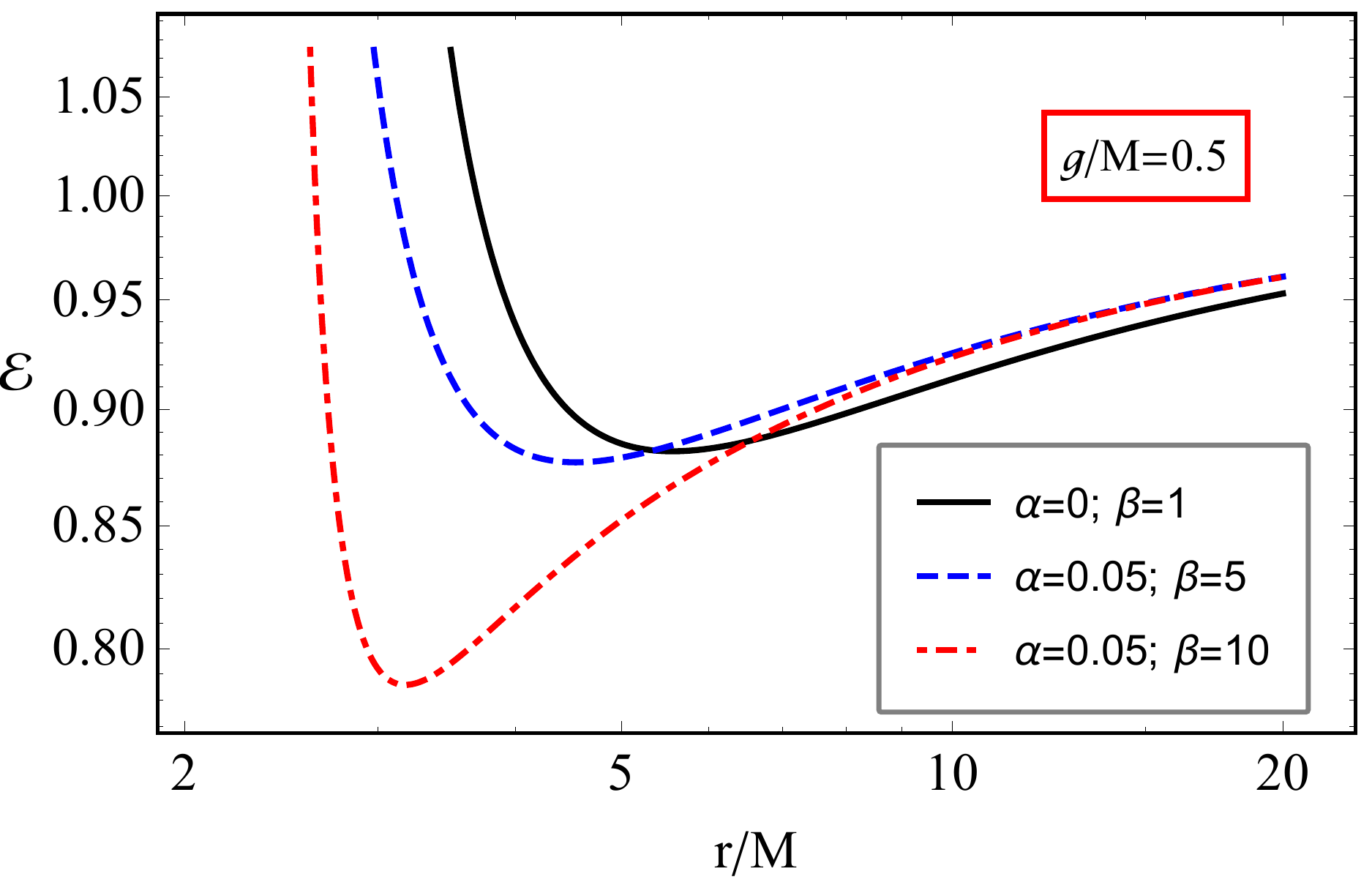}
\includegraphics[width=0.97\linewidth]{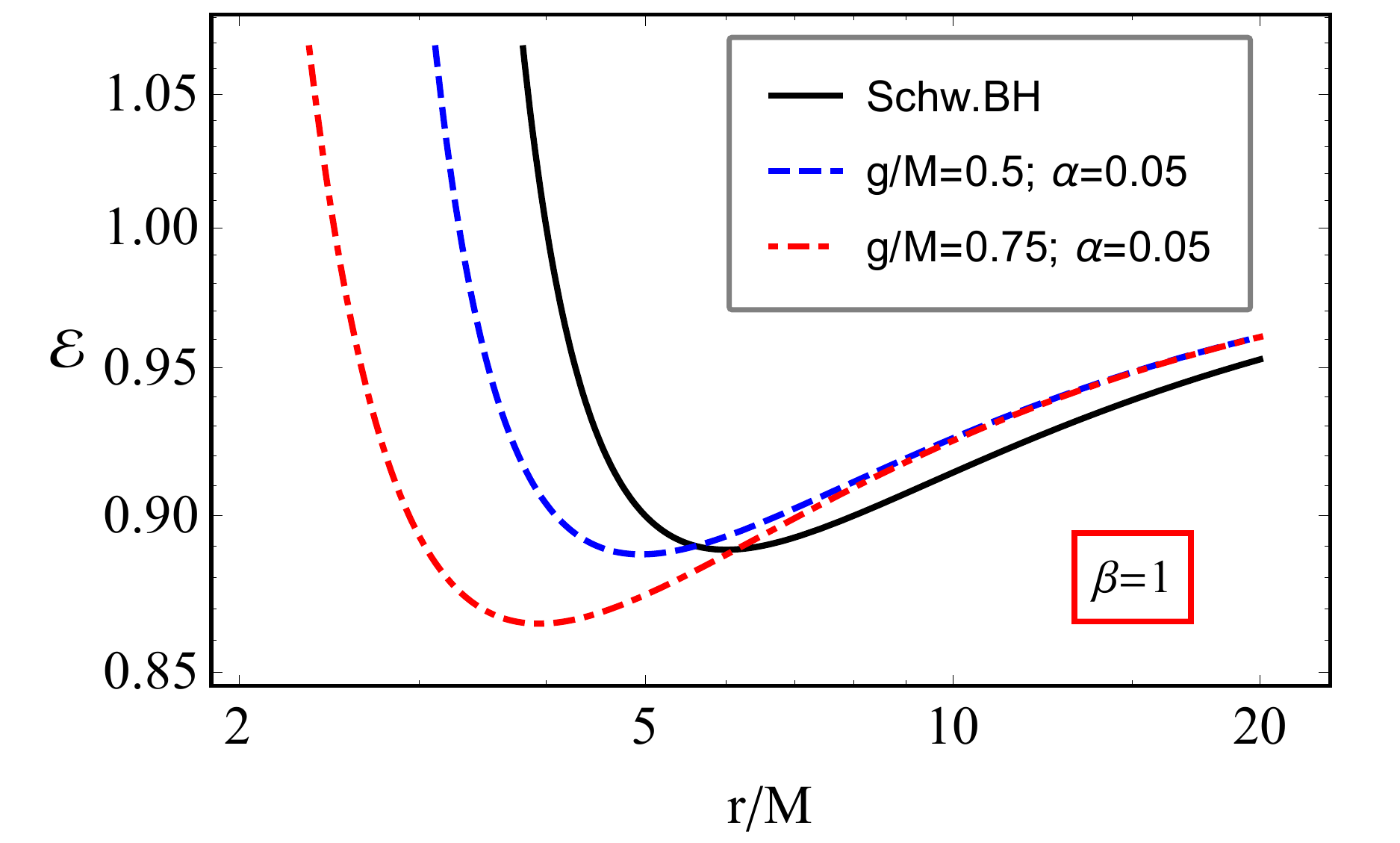} \caption{Specific energy of the magnetic dipole around magnetically charged regular Bardeen black hole in the presence of the perfect dark matter fluid in the circular orbits for the different values of magnetic charge parameter $g$, the dark matter parameter $\alpha$ and the parameter $\beta$. \label{EEBardeen}}\end{figure}

 Figure~\ref{EEBardeen} illustrates the radial dependence of the specific energy of the magnetic dipole for the fixed values of the magnetic charge parameter $g/M=0.5$ (on the top panel) and the parameter $\beta=1$ and the different values of the dark matter parameter. One can see from the top panel of the figure that the increase of both dark matter parameter and the parameter $\beta$ causes to decrease the specific energy of the magnetic dipole in circular orbits. Moreover, it is seen from the bottom panel of Fig.\ref{EEBardeen} that the specific energy decreases with the increase of the magnetic charge parameter of regular Bardeen black hole in the presence of the perfect fluid dark matter .
Here we provide analyses of effects of the dark matter parameter, magnetic charge of the Bardeen regular black hole and the parameter $\beta $ on the ISCO profiles presenting them in plot form based on numeric calculations.
\begin{figure}[ht!]\centering
\includegraphics[width=0.9\linewidth]{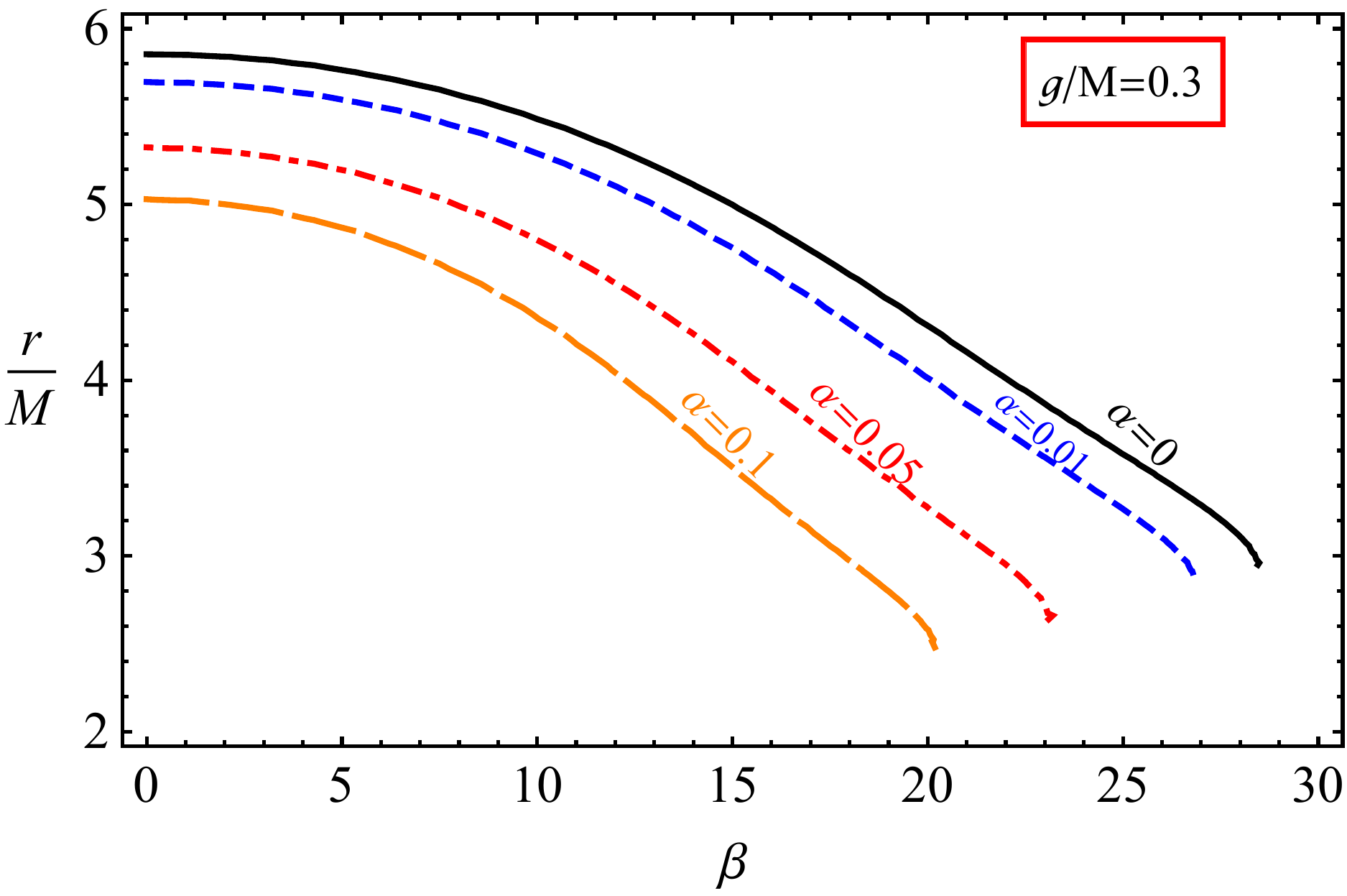}
\includegraphics[width=0.9\linewidth]{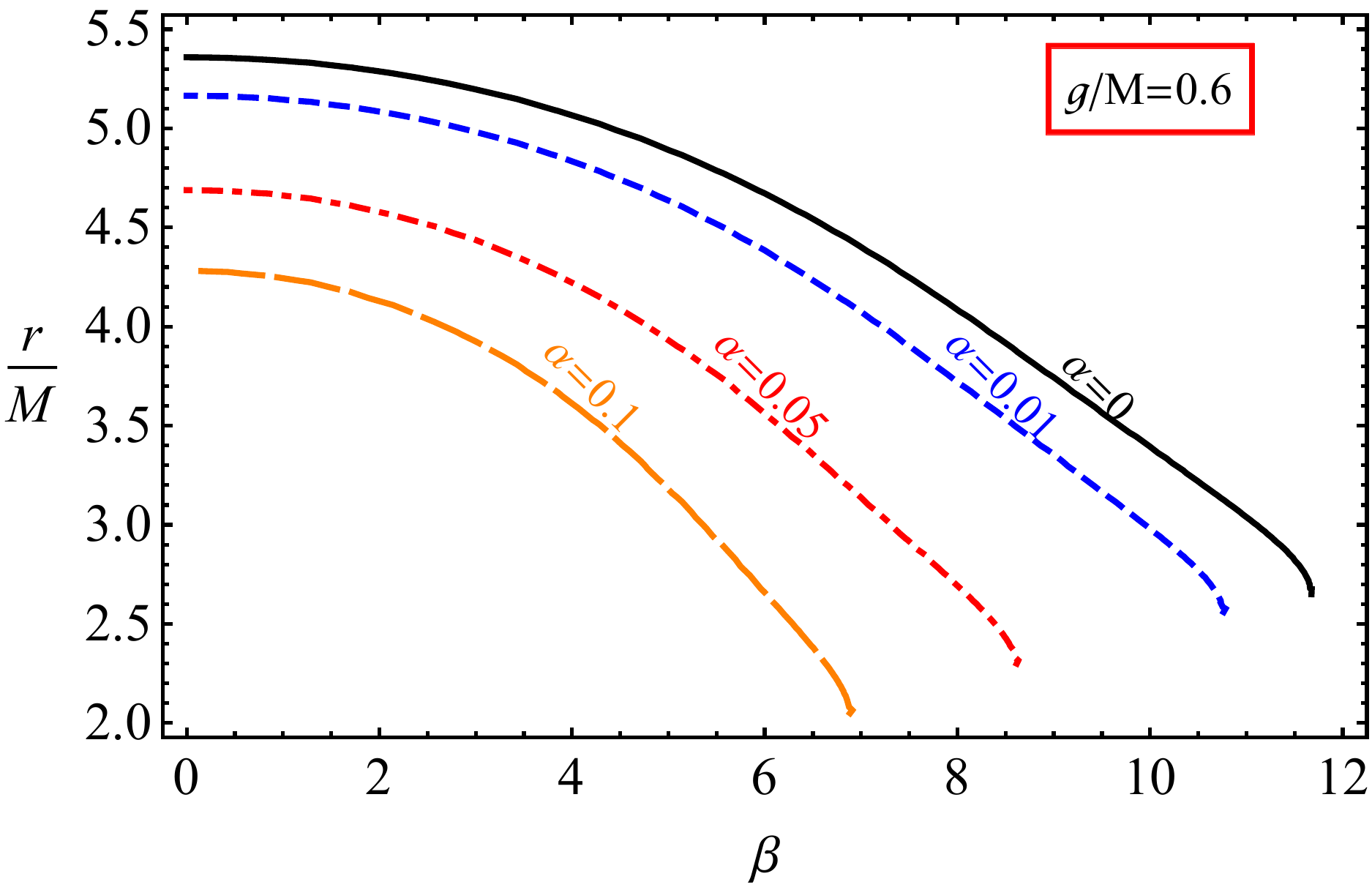}
 \caption{The dependence of ISCO radius of magnetic dipole around the magnetically charged Bardeen black hole in a perfect dark matter fluid from the parameter $\beta$ for fixed values of the magnetic charge ($g=0.3M$ on the top panel and $g=0.6M$ on the bottom one) for the different values of the dark matter parameter $\alpha$. \label{ISCOBardeen}}\end{figure}

ISCO radius of magnetic dipole around regular Bardeen black holes in the presence of the perfect fluid dark matter is demonstrated in Fig.\ref{ISCOBardeen} for the different values of the parameter $\beta$. One may see that at higher values of the parameter $\beta$ the ISCO does not exist due to dominance of magnetic interaction which plays a repulsive role  or attractive forces depending on the orientation of the magnetic dipole moment on the equatorial plane being parallel to the magnetic field generated by magnetic charge of regular Bardeen black hole. The upper limit for the parameter $\beta$, which is responsible for the existence of ISCO radius, decreases with the increase of the magnetic charge and the dark matter intensity parameters.

\section{Astrophysical applications \label{application}}

One of the real astrophysical scenarios of the studies of magnetic dipole motion around a black hole with magnetic interaction is possibility to probe black holes using  magnetized neutron stars dynamics around supermassive black holes. However, by now, there is no neutron star observed as recycled pulsars around supermassive black hole Sagittarius A* due to scattering of radio waves in the plasma medium surrounding SgrA* either circular orbits are not allowed for radio pulsars due to the destructive nature of magnetic interaction of magnetic moment of pulsar with ambient magnetic field. From this point of view, explorations of magnetized neutron stars motion around magnetically charged SMBHs is one of the most actual issues in relativistic astrophysics. In fact most astrophysical black holes are treated as (extremely) rotating Kerr black holes. However, the effects of spin of Kerr black holes on ISCO radius of test particles can be mimicked by other alternative gravity parameters and this may cause  undistinguishable results of analysis of observational data. Thus, here we concentrate on a simple analysis of degeneracy values of spin of Kerr black holes and magnetic charge of regular Bardeen black holes providing exactly the same values for ISCO of magnetic dipoles. Here we will treat the magnetar orbiting around SgrA* called PSR 1745-2900 as a magnetic dipole and we use the parameter calculated above for the system as $\beta= 10.2$. We would like to note that since the magnetic dipole behaves the same as a test neutral particle in the pure Kerr metric (with no external magnetic field) the ISCO radius of a magnetic dipoles orbiting around Kerr black holes can be expressed by the following well-known expression for the neutral ones in the case of  co- and contra-rotating ($\pm$ signs) motions~\cite{Bardeen72}
\begin{eqnarray}
r_{\rm isco}= 3 + Z_2 \pm \sqrt{(3- Z_1)(3+ Z_1 +2 Z_2 )} \ ,
\end{eqnarray}
where
\begin{eqnarray} \nonumber
Z_1 &  = & 
1+\left( \sqrt[3]{1+a}+ \sqrt[3]{1-a} \right) 
\sqrt[3]{1-a^2} \ ,
\\ \nonumber
Z_2^2 -Z_1^2 & = & 3 a^2 \ .
\end{eqnarray}

Now, we concentrate on possible cases to distinguish effects of magnetic charge of Bardeen regular black hole in the presence of the PFDM and rotating Kerr black holes on the investigations of the dynamics of magnetic dipole assuming that magnetized particle is the magnetar (SGR) PSR J1745-2900 orbiting around SgrA* with the parameters $\beta=10.2$.

 \begin{figure}[ht!]\centering\includegraphics[width=0.9\linewidth]{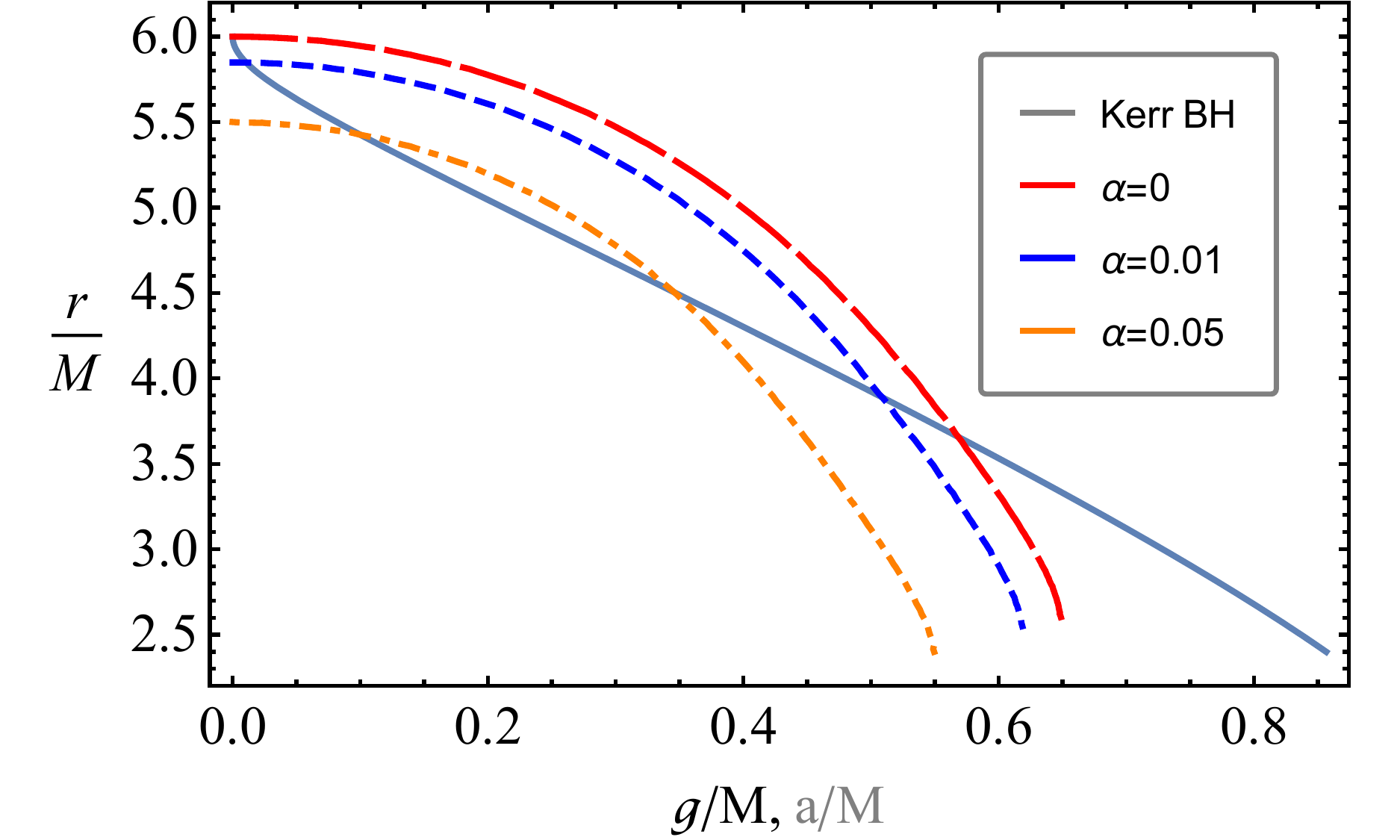}
 \includegraphics[width=0.85\linewidth]{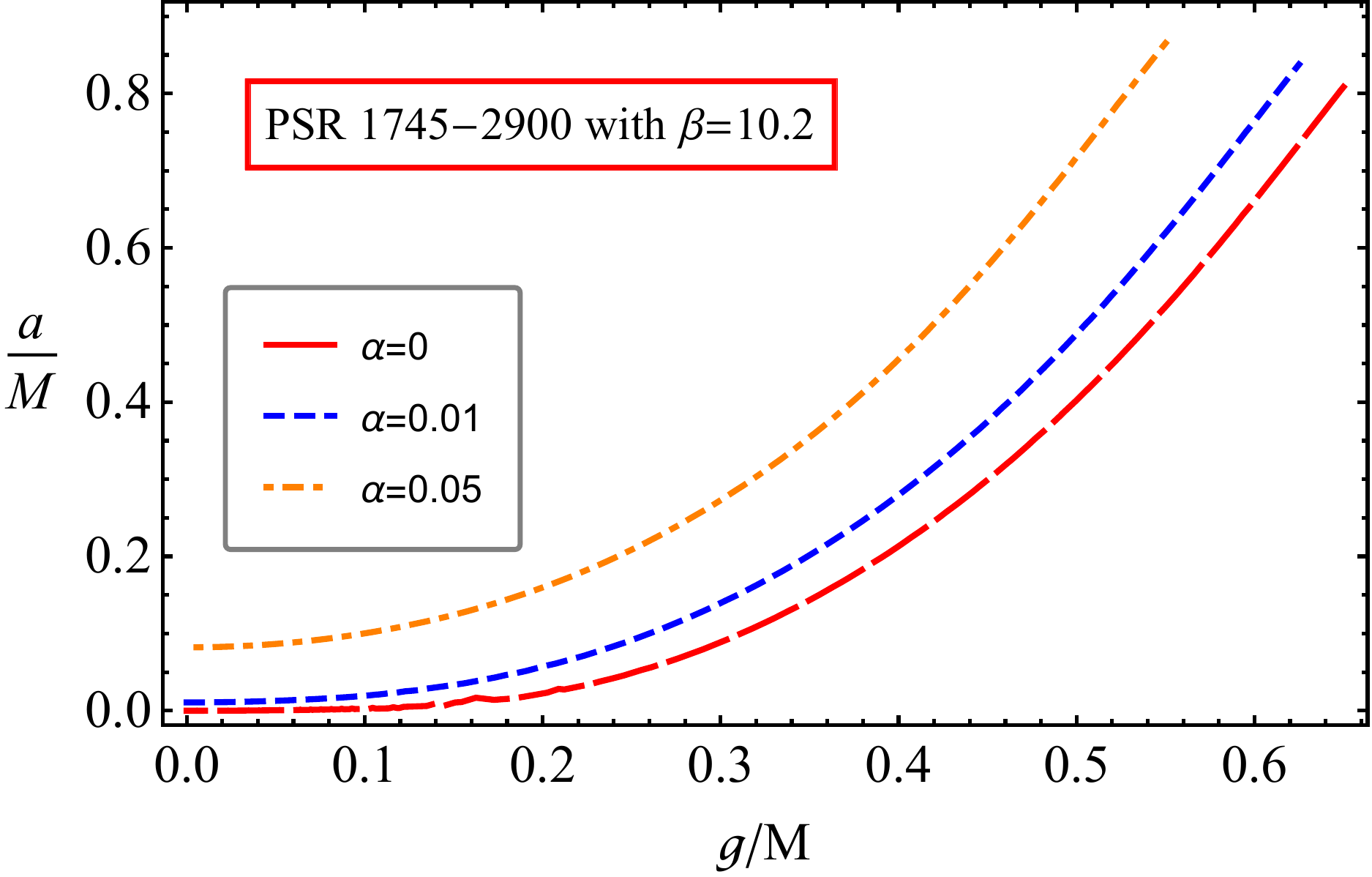}
 \caption{Dependence of ISCO radius of a magnetized particle with the parameter $\beta=10.2$ from magnetic charge of regular Bardeen black hole and the spin parameter of Kerr black hole for the different values of the dark matter parameter(on the top panel). Relation between magnetic charge of Bardeen regular black hole and spin of rotating Kerr black hole providing the same value of ISCO radius of the magnetic dipole with the parameter $\beta=10.2$ (on the bottom panel). \label{QvsQm}}\end{figure}

 In the top panel of Fig.\ref{QvsQm}, the dependence of magnetized particle's  ISCO radius on magnetic charge of regular Bardeen black hole for the different values of the dark matter parameter and spin of Kerr black hole is presented. In the bottom one, the degeneracy between dimensionless spin parameter of Kerr black hole and magnetic charge of Bardeen regular black hole for the different values of the dark matter intensity parameter is plotted. We have treated here the magnetar PSR 1745-2900 orbiting SgrA* as a magnetic dipole with the parameter $\beta=10.2$. One may easily see from the top panel that spin of Kerr black hole and magnetic charge of regular Bardeen black hole cause to decrease of ISCO radius of magnetic dipole. Our interest is in which degeneracy values dimensionless spin and magnetic charge parameters provide exactly the same values for ISCO radius of the magnetic dipole with the parameter $\beta=10.2$. Our numerical analysis based on the idea that two spacetime metrics provide the same ISCO indicates that in the absence of the perfect fluid dark matter the magnetic charge of the regular Bardeen black hole can mimic the spin of Kerr black hole in the range of $a/M\simeq (0\div 0.8085)$. 
 In this case the upper limit for the magnetic charge which provides the existence of ISCO is $g_{\rm upper}\simeq 0.65M$. Moreover, when $\alpha=0.01 (0.05)$ the upper limit for the magnetic charge decreases and equals to $g_{\rm upper}\simeq 0.62 M$ ($0.548M$) mimicking the spin parameter in the range of $a/M \in (0.0106 \div 0.8231 )$ ( $a/M \in (0.0816 \div 0.8595)$). It is seen that the mimic range of the spin parameter slightly shifts upward as the perfect fluid dark matter parameter increases.

In the absence of the PFDM, the ISCO radius of the magnetic dipole with the parameter $\beta=10.2$ has the values $r_{\rm isco} = 6M$ and $3.63M$. The first value corresponds to the classical Schwarzschild BH. The second value mimics the Kerr black hole with $a/M=0.5763$ or Bardeen BH with the value $g/M=0.5763$. In the existence of PFDM with the parameter $\alpha=0.01$ ($\alpha=0.05$), the ISCO radius takes exactly the same values at the two different values of the spin and magnetic charge parameter $a/M=g/M\simeq 0.0054$ and $a/M=g/M\simeq 0.5134$  ($a/M=g/M\simeq 0.3488$ and $a/M=g/M\simeq 0.1126$) providing the following ISCO radius $r_{\rm isco}= 5.89368M$ and $r_{\rm isco}=3.8714M$ ($r_{\rm isco}=4.4931M$ and $r_{\rm isco}= 5.3765M$), respectively. We would like to point out the importance of the cases when $a=g$ provides the same value for ISCO radius of the magnetic dipole and it is very challenging to distinguish the type of central black hole whether it is a rotating Kerr or magnetically charged Bardeen one through the analysis of the observational properties of an object with magnetic dipole such as pulsars and magnetars around a supermassive black hole.

\section{Conclusion\label{conclusion}}

In this work as first step neutral particle motion around the Bardeen black hole surrounded by PFDM is investigated. From the equation of motion it has been shown how the magnetic charge of the black hole and the intensity parameter of surrounded PFDM can influence the ISCO radius of a test particle and based upon the results obtained it has also been shown that  these parameters together can almost completely mimic the rotation parameter of Kerr black hole. In turn it leads to the assumption that black holes in the Universe cannot only be explained by the Kerr metric that characterizes the rotating black holes but also might be static ones with extra parameters.

We then considered a magnetically charged test particle motion in the Bardeen regular black hole spacetime surrounded by perfect fluid dark matter. We found that the ISCO radius is strongly affected as a consequence of combined effect of magnetic charge parameter $\lambda$ and perfect fluid dark matter as well as black hole  magnetic charge parameter $g$. From an observational point of view, distant  observer cannot distinguish any two geometries by analyzing electromagnet radiation emitted by accretion disk around black hole. In this respect, the combined effect of black hole magnetic charge parameter and perfect fluid dark matter plays an important role in determining the geodesics of magnetically charged test particles around black hole and explains observable properties such as the ISCO of particles around black holes.  We have found that combined effect of perfect fluid dark matter and magnetic charge parameter can mimic black hole rotation parameter up to $a/M\approx 0.9$. 

Moreover, dynamics of magnetic dipole around magnetically charged regular Bardeen black hole in PFDM field have been also studied. It has been observed that ISCO for magnetic dipole disappears at the values exceeding the calculated upper value for the parameter $\beta$. The upper limit decreases with the increase of both the dark matter and magnetic charge parameters. Finally, as an astrophysical application, we have analyzed degeneracy effects of spin of Kerr black hole and magnetic charge of regular Bardeen black hole for the different values of the dark matter parameter providing exactly the same value for ISCO radius of a magnetized particle with the same value of the parameter $\beta=10.2$ of the magnetar called PSR J1745-2900 orbiting around Sgr A*. It has been observed that the magnetic charge of the pure regular Bardeen black hole can mimic the spin of Kerr black hole  up to $a/M \simeq 0.8085$, while upper limit for the magnetic charge which may provide ISCO for the magnetic dipole is $g_{\rm upper}\simeq 0.65M$. In the presence of PFDM with the parameter $\alpha=0.01 (0.05)$ the upper limit for the magnetic charge parameter decreases and equals to $g_{\rm upper}\simeq 0.62 M$ ($0.548M$) and consequently mimicker value for the spin parameter of Kerr black hole lies in the range of $a/M \in (0.0106 \div 0.8231 )$ ( $a/M \in (0.0816 \div 0.8595)$).

\section*{Acknowledgement}

This work was supported by the Innovation Program of the Shanghai Municipal Education Commission, Grant No.~2019-01-07-00-07-E00035, the National Natural Science Foundation of China (NSFC), Grant No.~11973019, and the Uzbekistan Ministry for Innovative Development, Grants No. VA-FA-F-2-008 and No. MRB-AN-2019-29. 
JR, AA and BA thank Silesian University in Opava for the hospitality during their visit. The research work of AA is supported by PIFI fund of the Chinese Academy of Sciences. BN acknowledges support from the China Scholarship Council (CSC), Grant No.~2018DFH009013.
\bibliographystyle{apsrev4-1}
\bibliography{gravreferences}

\end{document}